\documentclass{aa}
\usepackage{cleveref}
\usepackage{natbib}
\bibpunct{(}{)}{;}{a}{}{,} 
\usepackage{graphicx}
\usepackage{txfonts}
\usepackage{soul}
\usepackage{color, xcolor}
\soulregister\cite7
\begin{document}

   \title{Does a long-lived remnant neutron star exist after short gamma-ray burst GRB\,160821B?}

   \subtitle{}

   \author{Guang-Lei Wu
          \inst{1,2}
           \and
            Yun-Wei Yu\inst{1,2}
         \and
           Jin-Ping Zhu\inst{3}
            }

   \institute{Institute of Astrophysics,
   			  Central China Normal University, Wuhan 430079, China\\
              \email{ yuyw@mail.ccnu.edu.cn}
              \and
              Key Laboratory of Quark and Lepton Physics (Central
China Normal University), Ministry of Education, Wuhan 430079,
China
         \and
         	Department of Astronomy, School of Physics, Peking University, Beijing 100871, China
             \\
             }

   \date{}


\abstract{Mergers of double neutron stars (DNSs) could lead to the formation of a long-lived massive remnant NS, which has been previously suggested to explain the AT\,2017gfo kilonova emission in the famous GW170817 event. For an NS-affected kilonova, it is expected that a non-thermal emission component can be contributed by a pulsar wind nebula (PWN), which results from the interaction of the wind from the remnant NS with the preceding merger ejecta. Then, the discovery of such a non-thermal PWN emission can provide an evidence for the existence of the remnant NS. Similar to GRB\,170817A, GRB\,160821B is also one of the nearest short gamma-ray bursts (SGRBs). A candidate kilonova is widely believed to appear in the ultraviolet-optical-infrared afterglows of GRB\,160821B. Here, by modeling the afterglow light curves and spectra of GRB\,160821B, we find that the invoking of a non-thermal PWN emission can indeed be well consistent with the observational data. This may indicate that the formation of a stable massive NS could be not rare in the DNS merger events and, thus, the equation of state of the post-merger NSs should be stiff enough.}


   \maketitle
%

\section{Introduction}
Mergers of double neutron stars (DNSs) and black hole-neutron star binaries are important sources of gravitational wave (GW) for aLIGO/Virgo and other ground-based GW detectors. After a merger, a pair of collimated relativistic jets could be launched to generate a short-duration gamma-ray burst (SGRB) through the internal dissipations in the jet \citep{Paczynski1986,Eichler1989,Narayan1992,Nakar2007}. The SGRB can be observed if the jet axis does not deviate from the Earth direction too much \citep{Rezzolla2011,Paschalidis2015}. The interaction of the jet with the ambient interstellar medium can drive an external shock (ES) to produce broad-band long-lasting afterglows \citep{Rees1992,Mszros1997,Sari1998,Chevalier2000,Granot2002}. Accompanying with the jet launching, a non-relativistic mass of $\sim10^{-4}-10^{-2}M_\odot$ can be ejected more widely due to the effects of tidal disruption, collision squeeze, and accretion feedback. It is suggested that nearly half of the elements heavier than iron in the universe can be synthesized in these neutron-rich merger ejecta, through the rapid neutron-capture process \citep[r-process;][]{Lattimer1974,Lattimer1976,Symbalisty1982}. Then, the radioactive decays of the $r$-process elements can effectively heat the ejecta to generate a bright thermal emission, which was first predicted by \cite{Li1998} and subsequently by \cite{Metzger2010}. This transient thermal emission is now usually termed as ``kilonova", since its peak luminosity is expected to be a few thousand times of that of the typical nova phenomena \citep{Roberts2011,Barnes2013,Kasen2013,Kasen2015,Kasen2017,Tanaka2013,Yu2013,Grossman2014,Metzger2014a,Metzger2014b,Wanajo2014,Perego2014,Martin2015,Li2016,Metzger2017,zhu2020}.

The first multi-messenger GW event had been discovered on 17 August, 2017. About 1.7\,s after the GW170817 signal detected by LIGO and Virgo \citep{Abbott2017a}, the \textit{Fermi} Gamma-ray Burst Monitor was successfully triggered by GRB\,170817A \citep{Abbott2017b,Goldstein2017,Zhang2018} and, subsequently, a large number of follow-up observations monitored the afterglow emission in different electromagnetic bands from the radio to X-rays \citep{Alexander2017,Hallinan2017,Margutti2017,Troja2017,D'Avanzo2018,Ghirlanda2019,Lazzati2018,Lyman2018,Ghirlanda2019} and as well as the kilonova AT\,2017gfo in the ultraviolet-optical-infrared band \citep{Abbott2017c,Andreoni2017,Arcavi2017,Chornock2017,Coulter2017,Covino2017,Cowperthwaite2017,Evans2017,Hu2017,Kilpatrick2017,Lipunov2017,Nicholl2017,Smartt2017,Soares-Santos2017,Tanvir2017}. The observations of GRB\,170817A and its afterglows robustly confirmed the long-standing hypothesis that SGRBs can originate from compact binary mergers. Moreover, it became possible to explore the angular structure of the SGRB jet from an off-axis view. Meanwhile, the observations of AT\,2017gfo indicated the existence of the merger ejecta, which suggests that the progenitor binary should at least contain one NS. In more detail, the existence of a ``blue" and maybe also a ``purple" component in the AT\,2017gfo emission further indicated that the merger product of the GW170817 event is very likely to be a hypermassive NS, which at least lasted for a few hundreds of milliseconds, since an immediately-formed black hole can only be associated with a ``red" kilonova\footnote{The reason of this judgement is that the neutrino emission from the remnant NS can suppress the synthesization of lanthanides in a part of merger ejecta and reduce its opacity.} \citep{Cowperthwaite2017,Perego2017,Tanaka2017,Tanvir2017,Villar2017,Kawaguchi2018}. Therefore, in summary, the progenitor of the GW170817 event can be identified as a DNS system, which is consistent with the result of the GW analysis.

However, strictly speaking, in the radioactive power model, the observationally required mass and opacity of the merger ejecta actually cannot fall in an acceptable parameter region predicted by the merger simulations. Therefore, alternatively, \cite{Yu2018} and \cite{Li2018} had modeled the AT\,2017gfo emission by invoking a long-lived post-merger NS. Such a remnant NS can naturally provide an extra energy source for the kilonova emission and thus reduce the requirement on the ejecta mass. Simultaneously, the remnant NS can also influence the opacity of the merger ejecta, due to the possible ionization of lanthanides by the hard emission from the NS. About 155\,days after GRB\,170817A, \cite{Piro2019} discovered a possible X-ray flare, which somewhat implied independently the existence of the remnant NS. Furthermore, by considering of the interaction between the merger ejecta and the relativistic wind from the remnant NS, it is expected that a non-thermal emission component could be generated by the shocked NS wind (i.e., the pulsar wind nebula; PWN), in addition to the thermal kilonova emission from the merger ejecta \citep{Kotera2013,Yu2019b}. Very encouragingly, such a non-thermal emission component had indeed been resolved from the AT\,2017gfo data, as presented in \cite{Ren2019}, which can improve the fitting to the AT\,2017gfo data significantly.

In view of the ultrahigh mass of the merger product around $2.5\,M_{\odot}$, it is undoubtedly necessary and important to further test the existence of post-merger NSs, which can provide a
robust constraint on the equation of state of the NS matter and then promote our understanding of the low-energy feature of strong interaction.
Besides the GRB\,170817A/AT\,2017gfo event, searchings for possible kilonova emission have already been implemented in the afterglows of many SGRBs since 2013 \citep{Berger2013,Tanvir2013,Yang2015,Jin2015,Jin2016,Jin2018,Jin2020,Gao2015,Gao2017,Kasliwal2017}. Among the SGRBs owning a kilonova candidate, GRB\,160821B is one of the lowest redshift of $z=0.162$. From its optical/nIR afterglow, an obvious excess was found. Because of its near distance, the kilonova emission associated with GRB\,160821B is in principle detectable and can provide a natural explanation for the observed optical/nIR excess \citep{Lamb2019,Troja2019}. In view of its luminosity lower than AT\,2017gfo, the kilonova after GRB\,160821B can in principle be modeled with a pure radioactive power. However, it could still be necessary to mention that a significant internal plateau had appeared in the early X-ray afterglow during the first few hundreds of seconds (see the insert in Figure \ref{LC}), which indicated that a post-merger NS also exists in this event. According to these observations, \cite{Ma2020} suggested that the post-merger NS could collapse into a black hole and then the subsequent kilonova could be powered by the accretion onto the black hole. Nevertheless, alternatively, as suggested by \cite{Yu2018}, the steep decay after the internal plateau may not represent the collapse of the NS, but just be caused by the suppression of the magnetic dipole radiation of the NS. In this case, the spin-down of the NS of a relatively low luminosity can still power the kilonova emission, which can be generally called as mergernovae \citep{Yu2013}. This scenario can provide a natural explanation for the AT\,2017gfo emission. Therefore, in our opinion, this situation could also appear in the case of GRB\,160821B. Then, this paper is devoted to test whether there is a non-thermal emission component arising from the interaction between the NS wind and the merger ejecta, just as mentioned above for AT\,2017gfo.

\section{The Model}
As a result of the collision of a relativistic NS wind with a preceding ejecta, a termination shock (TS) can be formed in the wind to decelerate the wind material, while the ejecta can be heated by absorbing the radiation from the shocked wind. Such an interaction has been previously studied in some semi-analytical works for GRBs \citep{Dai2004,Yu2007}, superluminous supernovae \citep{Kotera2013}, mergernovae \citep{Ren2019}, and even accretion-induced collapses of white dwarfs \citep{Yu2019b}. In this paper we employ the model proposed in \cite{Yu2019b} and \cite{Ren2019}, which is most relevant to the situation concerned here. I.e., a relativistic wind from a millisecond pulsar is blocked by a low-mass optically thick ejecta.
\subsection{The PWN emission}
The energy luminosity carried by a NS wind can usually be estimated by the luminosity of magnetic dipole radiation of the NS,
which reads
\begin{equation}\label{eq6}
	\begin{aligned}
		L_{\text{md}}&=\frac{B_{\text{p}}^{2}R_{\text{s}}^{6}}{6c^3}\left ( \frac{2\pi}{P_{}} \right) ^4=9.6\times 10^{42}B_{\text{p,}12}^{2}R_{\text{s,}6}^{6}P_{-3}^{-4}\text{erg\ s}^{-1}\\
	\end{aligned}
\end{equation}
with $B_{\text{p}}$, $R_{\text{s}}$, and $P_{}$ are the polar magnetic field strength, radius, and spin period of the NS, respectively, and $c$ is the speed of light. Hereafter, the conventional notation $Q_x = Q/10^x$ is adopted in cgs units. The temporal evolution of this wind luminosity is determined by the spin-down behavior of the NS, which can be written as
\begin{equation}\label{eq5}
	L_{\mathrm{md}} (t)=L_{\mathrm{md,i}}\left (1+\frac{t}{t_{\mathrm{sd}}}\right)^{-\alpha},
\end{equation}
where the initial value of the luminosity $L_{\rm{md,i}}$ is given for an initial spin period $P_{\rm i}$.
About the temporal index, we can take $\alpha=2$ when the spin-down is dominated by the magnetic dipole radiation. On the other hand, the NS's rotation could sometimes be braked primarily by a GW radiation, if the NS is deformed with a sufficiently high ellipticity $\epsilon$, which leads to $\alpha=1$. For these two different braking effects, the spin-down timescale of the NS can be expressed as
\begin{equation}\label{eq8}
t_{\text{sd,md}}=\frac{3Ic^3}{B_{\text{p}}^{2}R_{\text{s}}^{6}}\left (\frac{2\pi}{P_{\text{i}}}\right)^{-2}=2\times 10^3I_{45}R_{\text{s,}6}^{-6}B_{\text{p,}15}^{-2}P_{\rm i,-3}^{2}\,\text{s}
\end{equation}
and
\begin{equation}\label{eq7}
t_{\text{sd,gw}}=\frac{5P_{\text{i}}^{4}c^5}{2048\pi ^4GI\epsilon ^2 }=9.1\times 10^5\epsilon _{-4}^{-2}I_{45}^{-1}P_{\rm i,-3}^{4}\,\text{\rm s},
\end{equation}
respectively, where $ I $ is the inertia moment and $ G $ is the gravitational constant.

When the relativistic wind drives a TS by colliding with the preceding merger ejecta, a PWN (i.e., the shocked wind region) can be formed between the TS and the merger ejecta.  Denoting the bulk Lorentz factor of the unshocked wind by $\Gamma_{\rm w}$, the internal energy density of the PWN can be expressed by according to the shock jump condition
\begin{equation}\label{jump}
	e_{\rm ts}=4{\Gamma'}_{\rm ts}^2n'_{\rm w}m_{\rm e}c^2={\xi L_{\rm md}\over 4\pi R_{\rm ts}^2c},
\end{equation}
where $\Gamma'_{\rm ts}=\Gamma_{\rm w}/2$ is the Lorentz factor of the TS measured in the rest frame of the injecting unshocked wind, $n'_{\rm w}$ is the comoving number density of the wind electron/positrons, $ m_{\rm e} $ is the electron rest mass, $R_{\rm ts}$ is the radius of the TS, and the fraction $\xi$ is introduced due to a possible fact that the energy released from the NS is collimated in the jet direction. In our calculations, the value of $e_{\rm ts}$ is actually obtained according to the mechanical equilibrium between the PWN and the merger ejecta at the contact discontinuity surface (see Eq. \ref{equilibrium}). Then, by using Eq. (\ref{jump}), we can reversely obtain the evolution of the TS radius, which is determined by the motion of the inner boundary of the merger ejecta,

The electrons and positrons in the PWN can initially be accelerated by the TS to distribute with their random Lorentz factors as $dN_{\rm e}/d\gamma\propto \gamma^{-p}$ for $\gamma\geqslant\gamma_{\rm m}=[(p-2)/(p-1)]\Gamma'_{\rm ts}$, where $p$ is a constant spectral index. At a time of $t$, the total number of the accelerated electrons can be estimated by
\begin{equation}
	N_{\rm e}\approx \dot{N}_{\rm e}t={\xi L_{\rm md}t\over \Gamma_{\rm w}m_{\rm e}c^2}.
\end{equation}
Here, we multiply $L_{\rm md}$ as a function of time to the time $t$ directly but do not integrate $L_{\rm md}$ over $t$, because the electrons accelerated at early times can be cooled to be non-relativistic very quickly via their synchrotron radiation and, sometimes, further via synchrotron self-Compton scattering. For calculating the synchrotron radiation, we estimate the stochastic magnetic field in the PWN by $B_{\rm ts}=(4\pi\epsilon_{\rm B}e_{\rm ts})^{1/2}$ with a magnetic equipartition factor $\epsilon_{\rm B}$. At the very beginning, the stochastic magnetic field in the PWN could be very high, which can lead the cooling timescale of relativistic electrons (i.e., $\gamma_{\rm }\gtrsim 2$) to be much shorter than the dynamical timescale. In this case, the number of the relativistic electrons that can contribute to the synchrotron radiation should be discounted by a fraction of $t_{\rm col}/t$, which yields
\begin{equation}
	N_{\rm e,rel}=N_{\rm e}\times\min[1,{t_{\rm col}\over t}],
\end{equation}
where $t_{\rm col}=3\pi m_{\rm e}c/\sigma_{\rm T}B_{\rm ts}^2$ is taken for $\gamma_{\rm }\sim 2$ and $\sigma_{\rm T}$ is the Thomson cross section.

Following \cite{Sari1998}, we can analytically calculate the synchrotron radiation spectrum of relativistic electrons by
\begin{equation}\label{Lpwn}
	L_{\nu}^{\rm pwn}=L_{\nu,\max}^{\rm pwn}\times\left\{
	\begin{array}{ll}
		\left({\nu\over\nu_{\rm l}}\right)^{1/3},~~~~~~~~~~~~~~~~~~~~~~\nu<\nu_{\rm l};\\
		\left({\nu\over\nu_{\rm l}}\right)^{-(q-1)/2},~~~~~~~~~~~~~~~~\nu_{\rm l}<\nu<\nu_{\rm h};\\
		\left({\nu_{\rm h}\over\nu_{\rm l}}\right)^{-(q-1)/2}\left({\nu\over\nu_{\rm h}}\right)^{-p/2},~~~~\nu_{\rm h}<\nu,\\
	\end{array}\right.
\end{equation}
where the peak luminosity is given by $L_{\nu,\max}^{\rm pwn}=N_{\rm e,rel}{m_{\rm e}c^2\sigma_{T}
		} {B}_{\rm ts}/(3q_{\rm e})$
with $q_{\rm e}$ is the electron charge. For the braking frequencies and the spectral index, we have $\nu_{\rm l}=\min[\nu_{\rm m},\nu_{\rm c}]$,
$\nu_{\rm h}=\max[\nu_{\rm m},\nu_{\rm c}]$, and $q=2$ for
$\nu_{c}<\nu_{m}$ and $q=p$ for $\nu_{c}>\nu_{m}$, where $\nu_{\rm m}={ q_{\rm e}{B}_{\rm ts}{\gamma}_{\rm m}^2/2\pi m_{\rm e}c}$,  $\nu_{\rm c}={ q_{\rm e}{B}_{\rm ts}{\gamma}_{\rm c}^2/2\pi m_{\rm e}c}$,
and the cooling Lorentz factor is defined as $\gamma_{\rm c} = \max[2, 6\pi m_ec/\sigma_{\rm T}B_{\rm ts}^2t]$.

\subsection{The kilonova emission}
For the merger ejecta, as usual, we take a power-law density profile as \citep{Nagakura2014}
\begin{equation}\label{eq2}
	\rho_{\mathrm{ej}} (R , t) =\frac{ (\delta-3) M_{\mathrm{ej}}}{4 \pi R_{\max }^{3}}\left[\left (\frac{R_{\min }}{R_{\max }}\right)^{3-\delta}-1\right]^{-1}\left (\frac{R}{R_{\max }}\right)^{-\delta}
\end{equation}
with a distribution index of $\delta $, where $ M_\mathrm{ej} $ is the total mass of the ejecta, $ R_{\min}$ and $ R_{\max} $ are the minimum and maximum radii, respectively. Since the internal energy of the ejecta is usually much smaller than its kinetic energy, we assume that the ejecta expands homologously. Therefore, the maximum and minimum ejecta radii for a given time $t$ can be expressed as $R_{\rm max} = v_{\rm max}t$ and $R_{\rm min} = v_{\rm min}t$, by invoking the maximum and minimum velocities.
Following \cite{Metzger2017}, we separate the merger ejecta into $n$ mass layers and denote the layers by the subscript $i=1,2,\cdot
\cdot\cdot,n$, where $i=1$ and $n$ represent the bottom and the head layers, respectively. Then, the evolution of the internal energy $E_{{\rm int,}i}$ of the $i-$th layer can be determined by the energy conservation law as \citep{Kasen2010}
\begin{equation}\label{Eint}
	{dE_{{\rm int,}i}\over dt}=L_{{\rm h},i}+m_{i}\dot{q}_{{\rm r},i}\eta_{\rm
		th}-{E_{{\rm int,}i}\over R_{i}}{dR_{i}\over dt}-L_{{\rm e},i} ,
\end{equation}
where $L_{{\rm h},i}$ is the heating rate due to the absorption of the PWN emission by the layer, $\dot{q}_{\rm r}$ is the radioactive power per unit mass, $\eta_{\rm th}$ is the thermalization efficiency of the radioactive power, $m_{i}$ and $R_{i}$ are the mass and radius of the layer, $L_{{\rm e},i}$ is the observed luminosity contributed by this layer. The specific expressions of the terms in Equation (\ref{Eint}) would be introduced as follows.

First of all, the energy injection rate from the PWN to the $i-$th layer can be calculated by
\begin{equation}\label{taui}
	L_{{\rm h},i}=\int L_{\nu}^{\rm pwn}e^{-\tau_{\nu,i}}\left(e^{\Delta \tau_{\nu,i}}-1\right)d\nu,
\end{equation}
where the value of $L_{\nu}^{\rm pwn}$ is given by Eq. (\ref{Lpwn}) and the optical depths are defined as
\begin{equation}\label{}
	\tau_{\nu,i}=\int_{R_{\min}}^{R_i}\kappa_{\nu}\rho_{\rm ej}(r)d r
\end{equation}
and
\begin{equation}
	\Delta\tau_{\nu,i}=\int_{R_{i-1}}^{R_i}\kappa_{\nu}\rho_{\rm ej}(r)d r.
\end{equation}
When $n\rightarrow\infty$ and thus $\Delta \tau_{i}\rightarrow0$, we can get $L_{\rm h}=\Sigma_{i=1}^{n} L_{{\rm h},i}=\int L_{\nu}^{\rm pwn}\left(1-e^{-\tau_{\nu, \rm tot}}\right)d\nu$ from Eq. (\ref{taui}), where $\tau_{\nu, \rm tot}$ is the optical depth of the whole ejecta for a given frequency.
In our calculations, Eq. (17) in \cite{Yu2019b} is adopted to describe the frequency-dependent opacity $\kappa_{\nu}$, which was obtained by fitting the
numerical results presented in Figure 8 of \cite{Kotera2013}. Secondly, the radioactive power per unit mass reads \citep{Korobkin2012}
\begin{equation}\label{eq9}
	\dot{q}_{\mathrm{r}}=4 \times 10^{18}\left[\frac{1}{2}-\frac{1}{\pi} \arctan \left (\frac{t-t_{0}}{\sigma}\right)\right]^{1.3} \operatorname{erg} \mathrm{s}^{-1} \mathrm{~g}^{-1}
\end{equation}
with $ t_0=1.3\,{\rm s} $ and $ \sigma=0.11\,{\rm s} $, and the thermalization efficiency is given by \citep{Barnes2016,Metzger2017arXiv}
\begin{equation}\label{eq10}
	\eta_{\mathrm{th}}=0.36\left[\exp \left (-0.56 t_{\mathrm{day}}\right)+\frac{\ln \left (1+0.34 t_{\mathrm{day}}^{0.74}\right)}{0.34 t_{\mathrm{day}}^{0.74}}\right]
\end{equation}
with $t_{\rm day} = t/{\rm day}$.
Thirdly, the luminosity of the thermal emission of the $i-$th layer can be determined by the radiative diffusion as
\begin{equation}\label{eq11}
	L_{{\rm e},i} =\frac{E_i}{\max \left[ R_i/c,t_{\text{d,}i} \right]}	,
\end{equation}
where the radiation diffusion timescale of the $ i $-th layer is
\begin{equation}\label{eq12}
	t_{\text{d,}i}=\frac{3\kappa_{\rm es}}{4\pi R_ic}\sum_{i'=i}^n{m}_{i'},
\end{equation}
where $\kappa_{\rm es}=0.2\rm ~cm^2~g^{-1}$ is the electron-scattering opacity since the ejecta thermal emission is mainly in the UV/optical bands.
Under this treatment, the propagation of the PWN emission in the ejecta and the radiative transfer of the ejecta heat can be generally described by a set of independent equations for different layers.

After the calculations of all layers, we can finally obtain the total luminosity of the ejecta thermal emission as
\begin{equation}
	L_{\rm e}=\sum_{i=1}^{n} L_{{\rm e},i},
\end{equation}
which corresponds to a black-body temperature of $T_{\mathrm{eff}}=\left ({L_{\rm e}}/{4 \pi \sigma R_{\rm ph}^{2}}\right)^{1 / 4}$,
where $\sigma$ is the Stephan-Boltzmann constant, and $R_{\rm ph}$ is the photosphere radius where the electron-scattering optical depth satisfies $(\tau_{\rm es, tot}-\tau_{{\rm es,}i})= 1$. If $R_{\rm ph} < R_{\rm min}$, we simply set $R_{\rm ph} = R_{\rm min}$. For an effective black-body spectrum, the luminosity of the kilonova emission at a frequency $\nu$ can be given by
\begin{equation}\label{eq15}
	L_\nu^{\rm kn}=  (1 + z) {8 \pi^2 R_{\rm{ph}}^{2}\over c^2} {h \nu'^{3}\over {\exp \left (h \nu' / k T_{\mathrm{eff}}\right)-1}},
\end{equation}
where $ h $ is the Planck constant, $ k $ is the Boltzmann constant, and $\nu' =  (1 + z)\nu$. Meanwhile, by using the internal energy of the innermost layer $E_{{\rm int,}1}$ and according to the the mechanical equilibrium between the PWN and the merger ejecta, we can obtain the internal energy density of the PWN by
\begin{equation}
e_{\rm ts}={E_{{\rm int,}1}\over 4\pi R_{\min}^2\Delta R},\label{equilibrium}
\end{equation}
where $\Delta R=(R_{\max}-R_{\min})/n$. Then, we can finally combine the calculations for the PWN emission and the kilonova together.

\begin{table*}[tbph]
	\caption{The adopted parameter values for the tentative fitting of the observations of GRB\,160821B} \label{table1}
	\begin{center}
		\begin{tabular}{lllllll}
			\hline \hline
			jet external shock
			\\ \hline
			$ E_{\rm{j}} $	& $ n_0$		& $ \epsilon_{\rm{e}} $	& $ \epsilon_{\rm{B}} $	& $ p $	& $ \theta_{\rm j} $	 \\
			$ 10^{50}\rm{erg} $	& $ 1.58\times10^{-3}\,\rm{cm^{-3}} $	& $ 0.5 $ 	& $ 0.006 $  	& $ 2.3 $	& $ 14^{\circ} $		 \\
			
			\hline \hline
			kilonova
			\\ \hline
			$\xi L_{\rm{md,i}}  $	& $t_{\rm sd} $	& $ M_{\rm{ej}}$	& $ \kappa_{\rm es}$ 		& $ v_{\rm{min}} $	& $ v_{\rm{max}} $	& $ \delta  $ \\  	
			
			$ 10^{41} \rm{erg \;s^{-1}} $	& $ 8.64\times10^5 \; \rm{s}$	& $ 0.01\, M_{\odot} $	& $ 0.2 \,\rm{g\;cm^{-2}} $	& $ 0.1 \,c $	& $ 0.4\,c $	 &$ 1.5 $	\\
			\hline \hline
			PWN
			\\ \hline
			$ \Gamma_{\rm{w}} $	&	$ \epsilon_{\rm{e}} $  &	$ \epsilon_{\rm{B}} $ &  $ p $	& $  $\\
			\hline
			$ 10^5 $ & $ 0.99 $	& $ 0.01 $	& $ 2.5	 $ \\
			\hline \hline
			\label{ta1}
		\end{tabular}%
	\end{center}
	\par
\end{table*}

\begin{figure*}[htpb]
    \centering
    \includegraphics[width = 0.32\linewidth , trim = 70 30 95 50, clip]{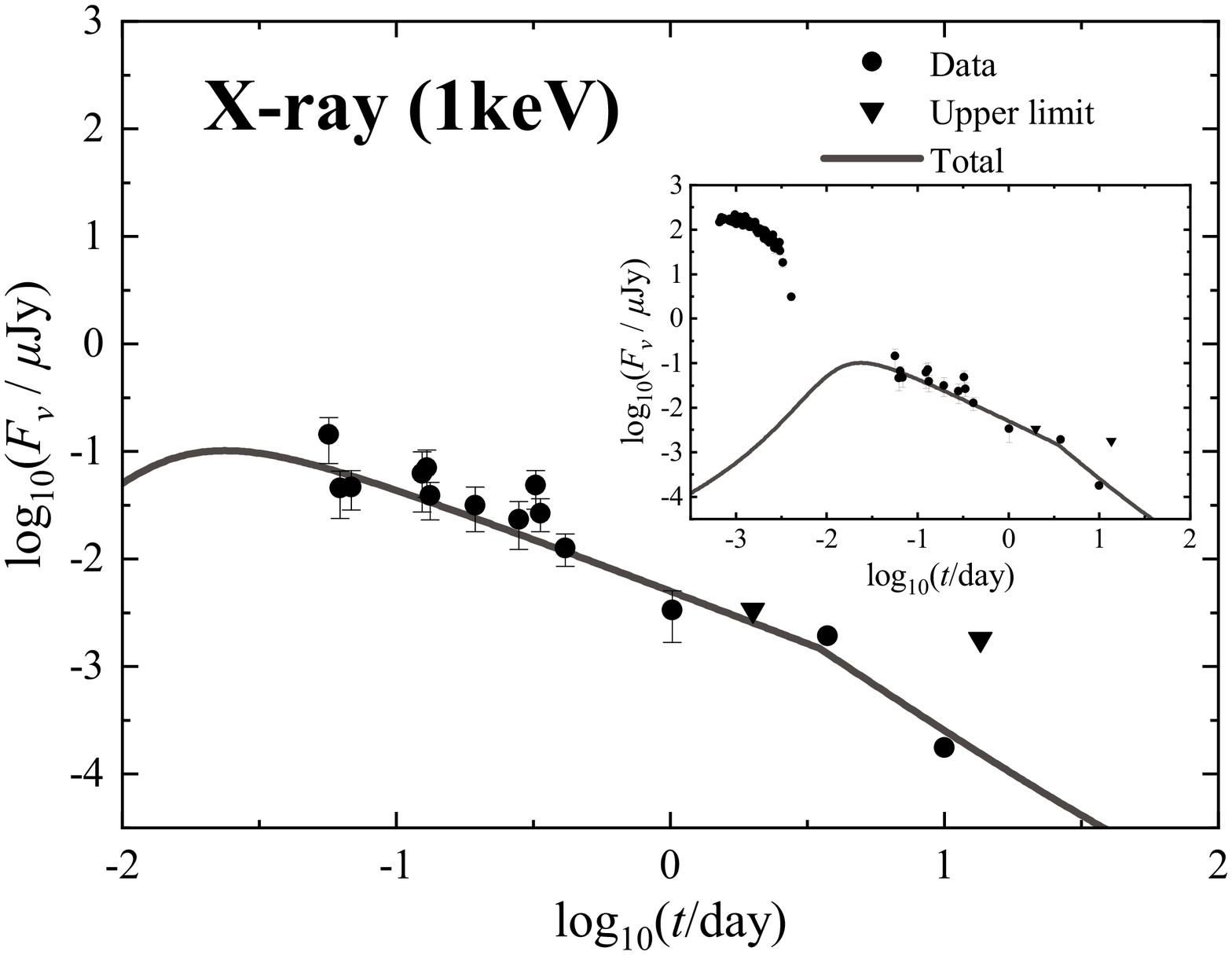}
    \includegraphics[width = 0.32\linewidth , trim = 70 30 95 50, clip]{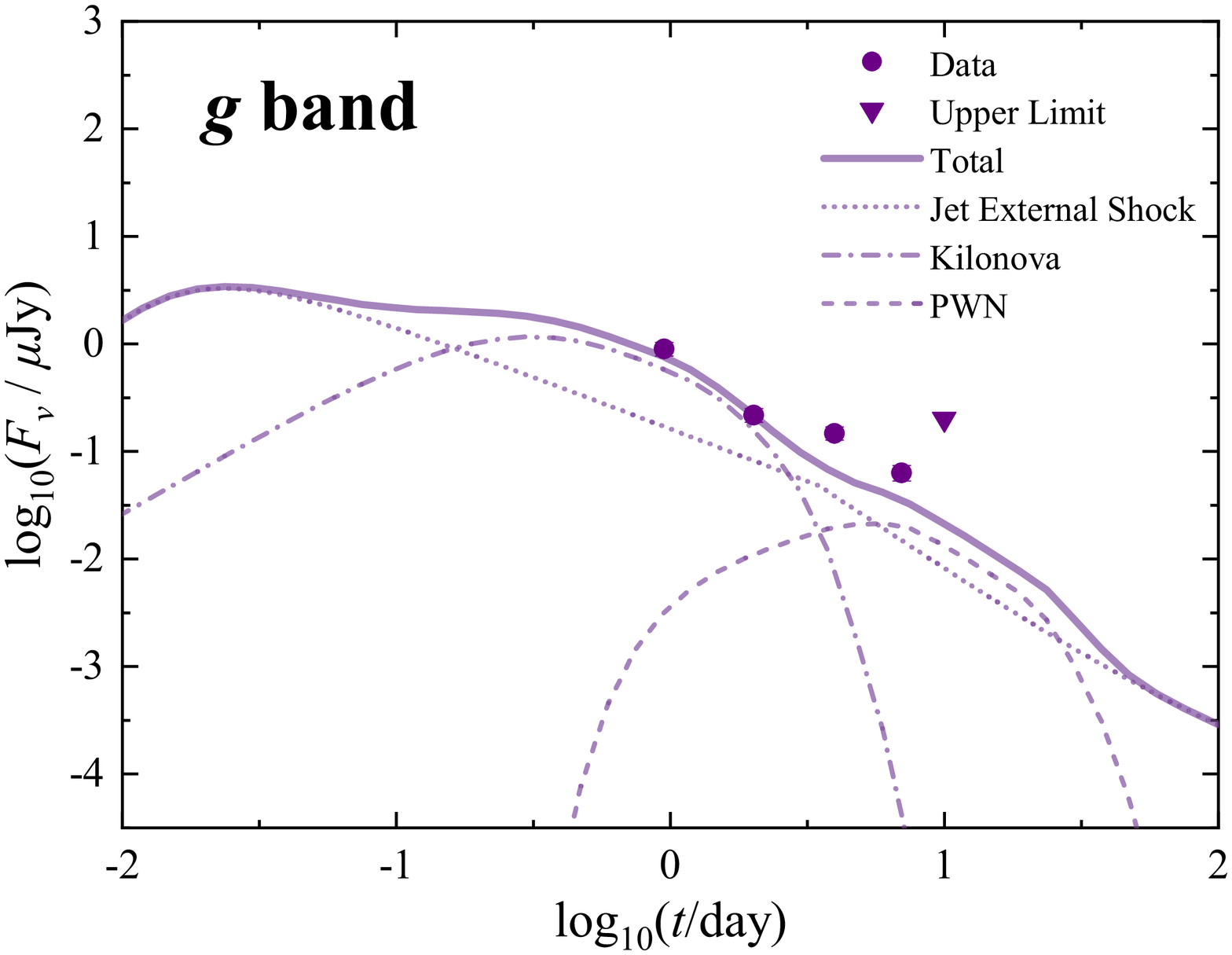}
    \includegraphics[width = 0.32\linewidth , trim = 70 30 95 50, clip]{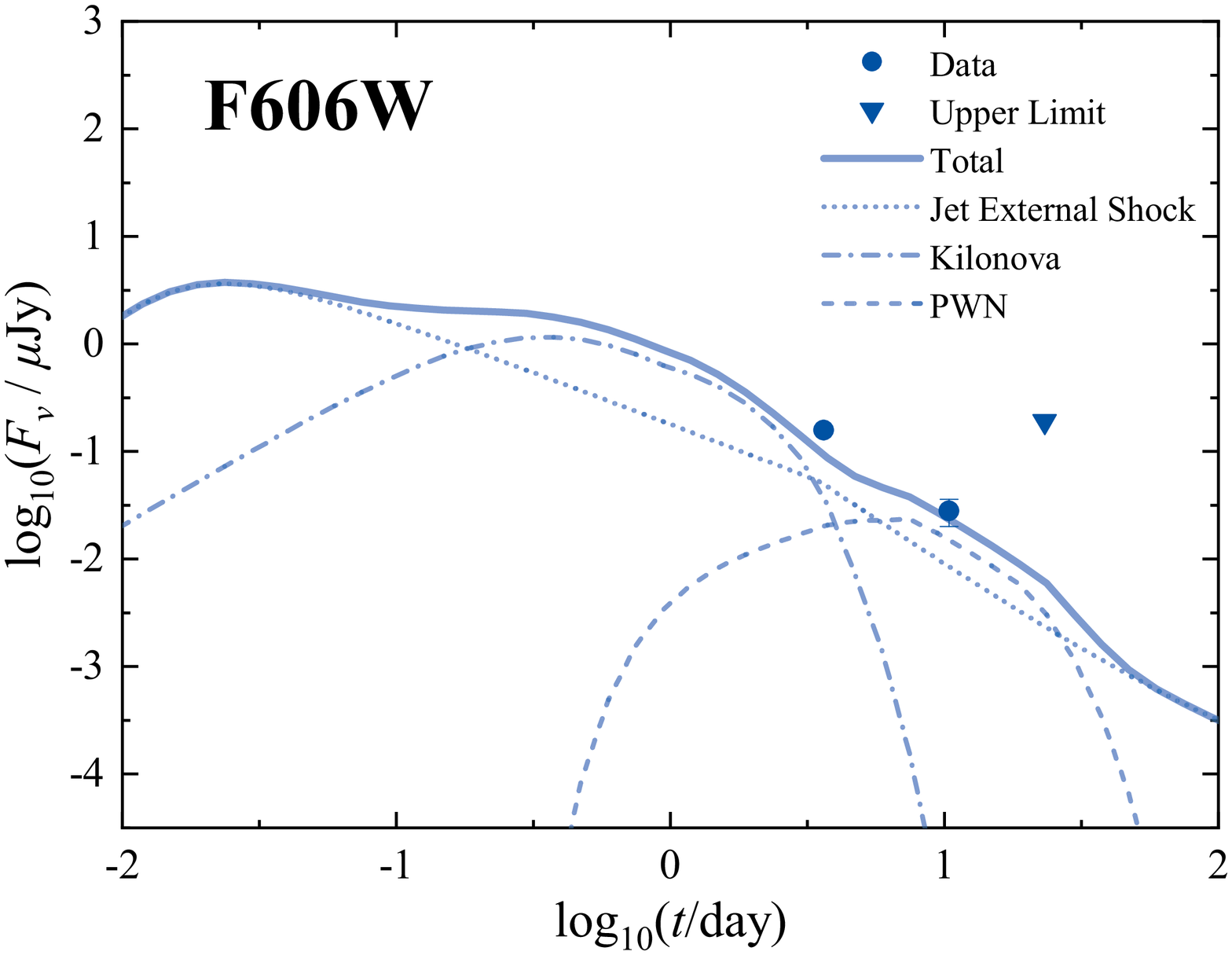}
    \includegraphics[width = 0.32\linewidth , trim = 70 30 95 50, clip]{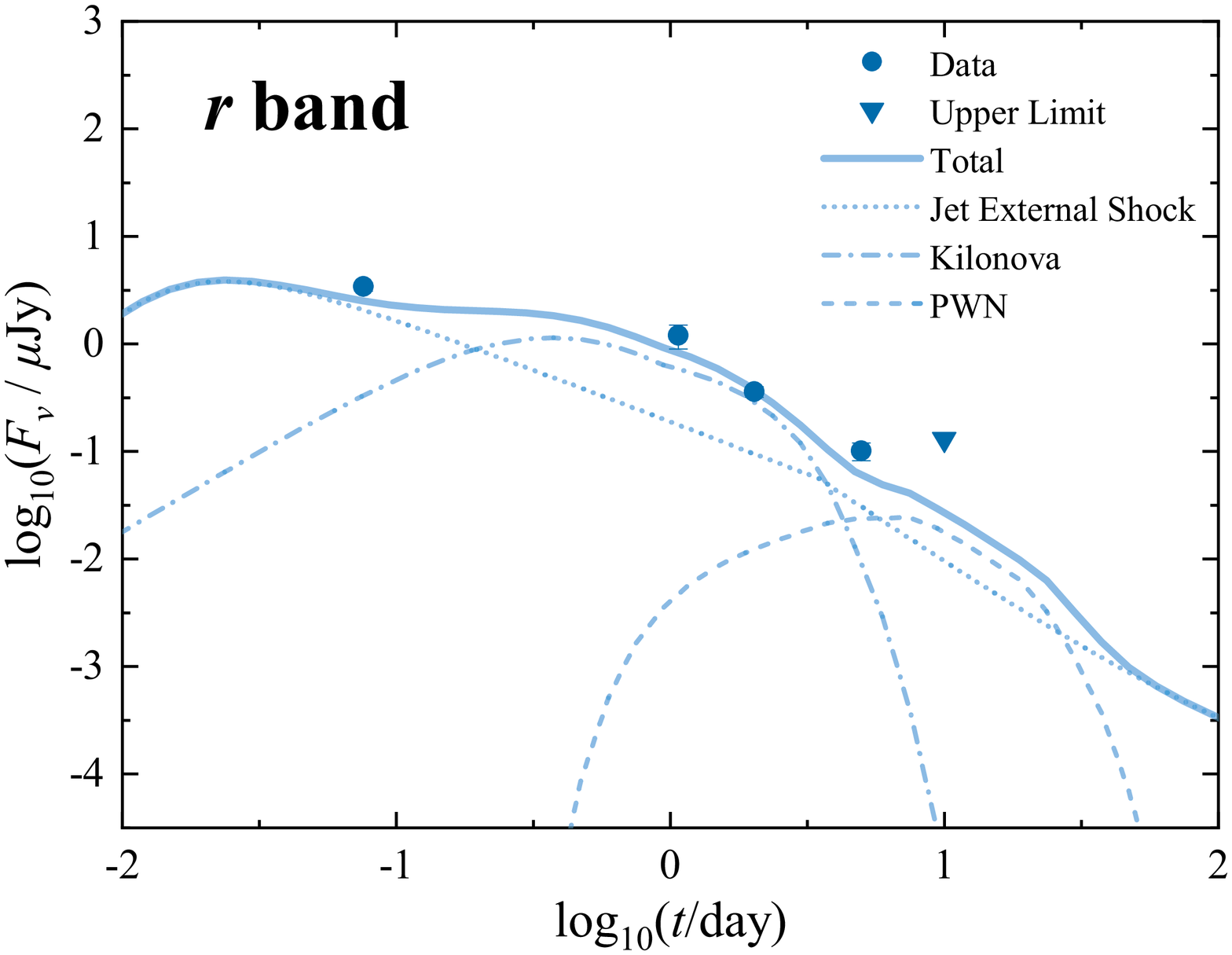}
    \includegraphics[width = 0.32\linewidth , trim = 70 30 95 50, clip]{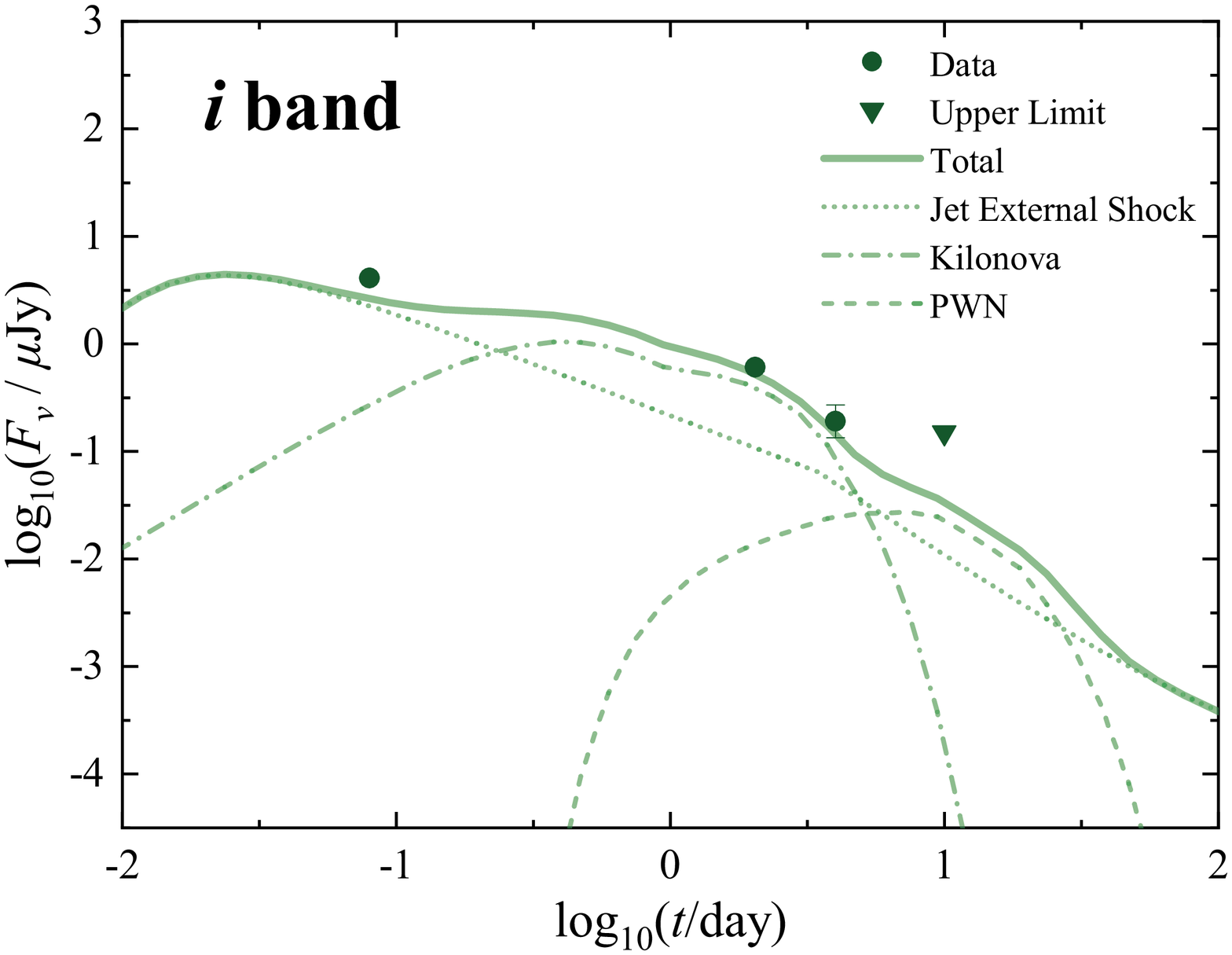}
    \includegraphics[width = 0.32\linewidth , trim = 70 30 95 50, clip]{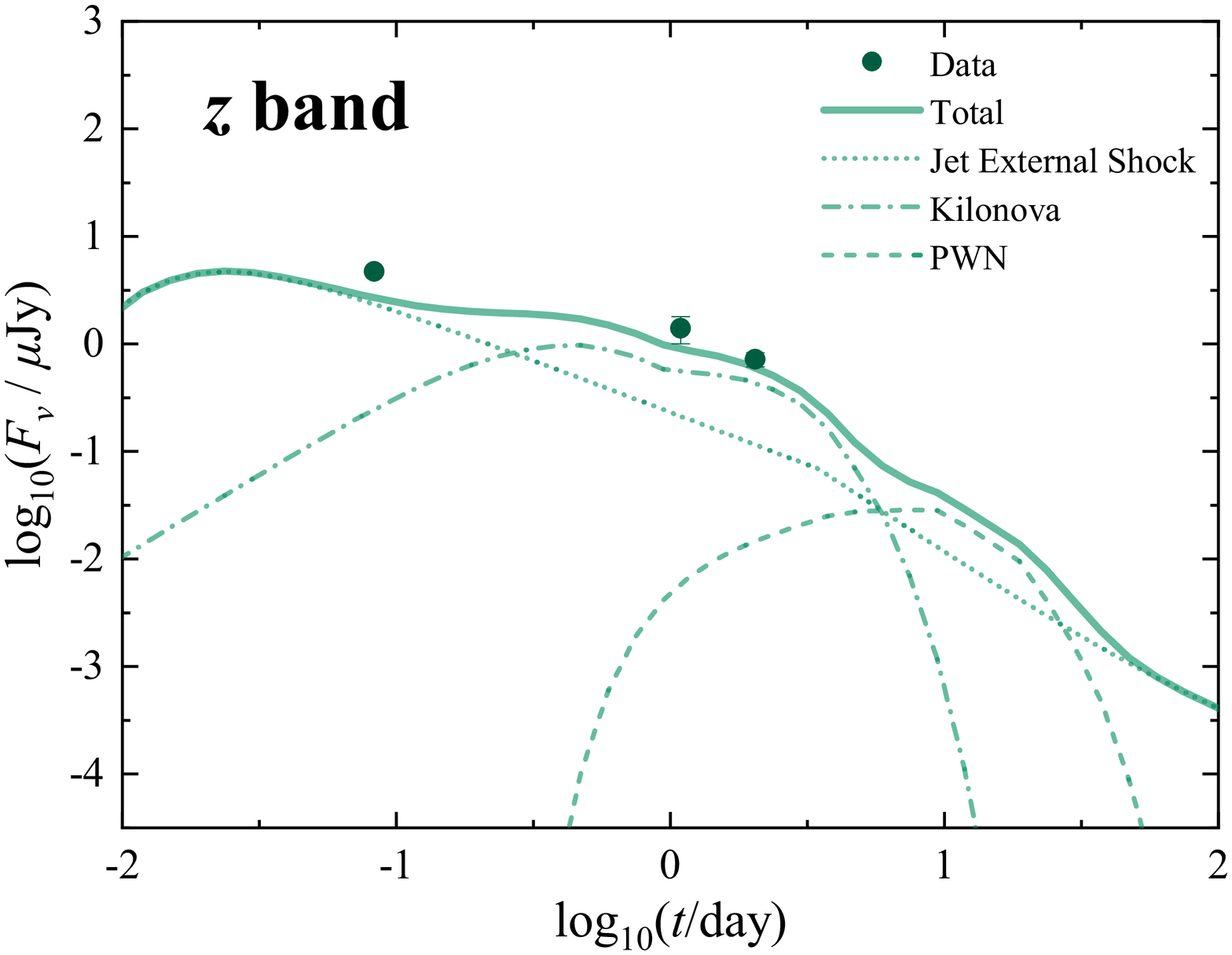}
    \includegraphics[width = 0.32\linewidth , trim = 70 30 95 50, clip]{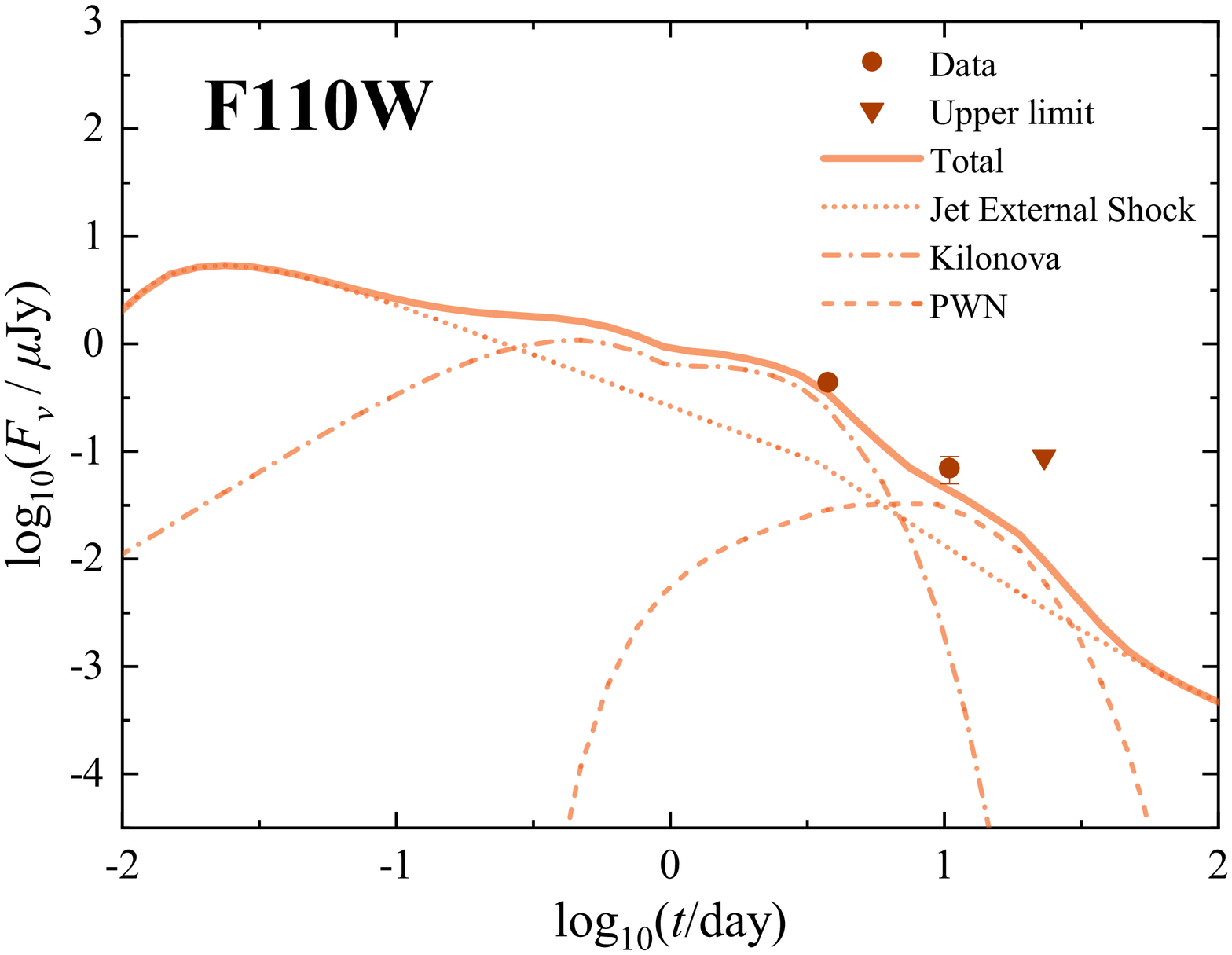}
    \includegraphics[width = 0.32\linewidth , trim = 70 30 95 50, clip]{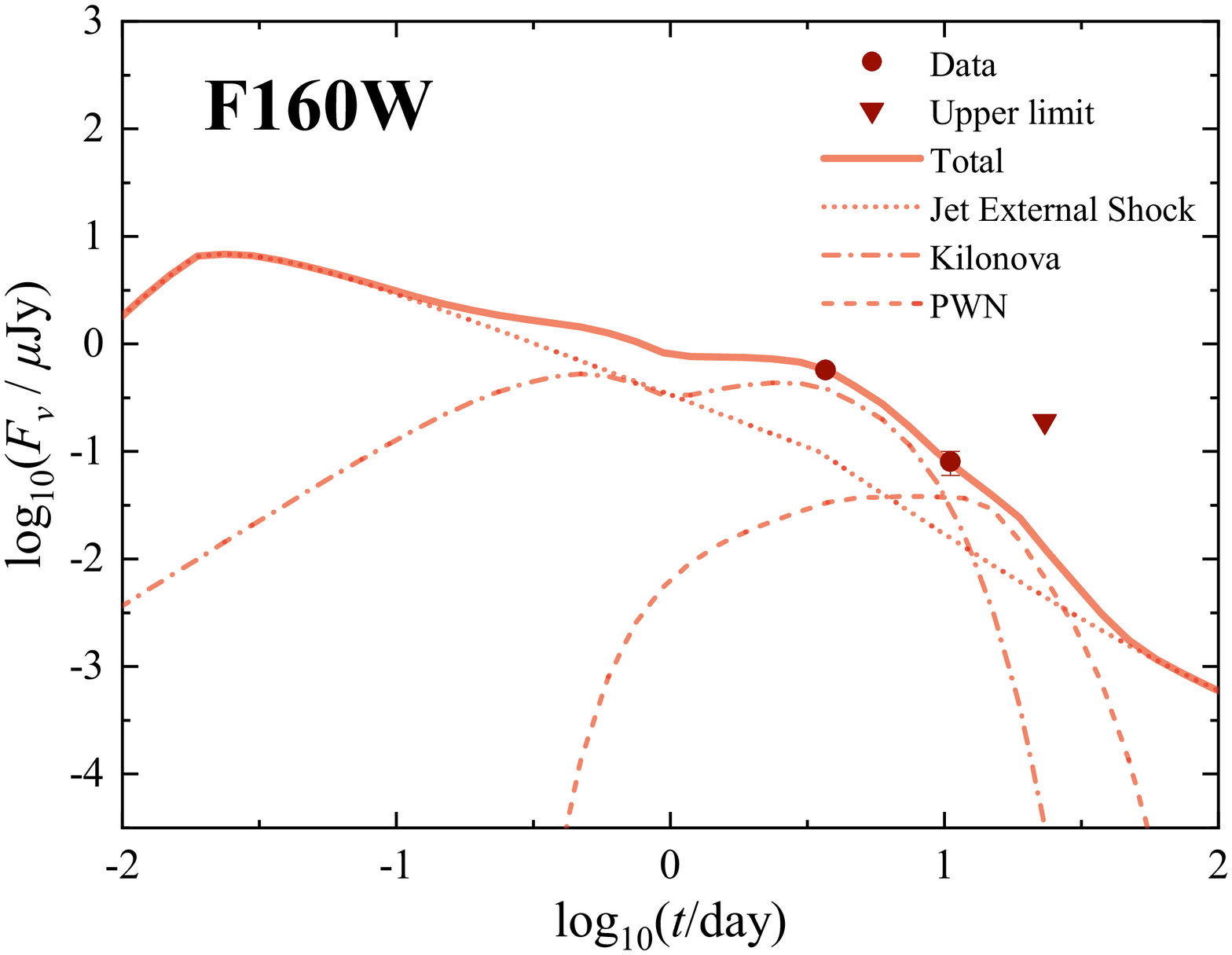}
    \includegraphics[width = 0.32\linewidth , trim = 70 30 95 50, clip]{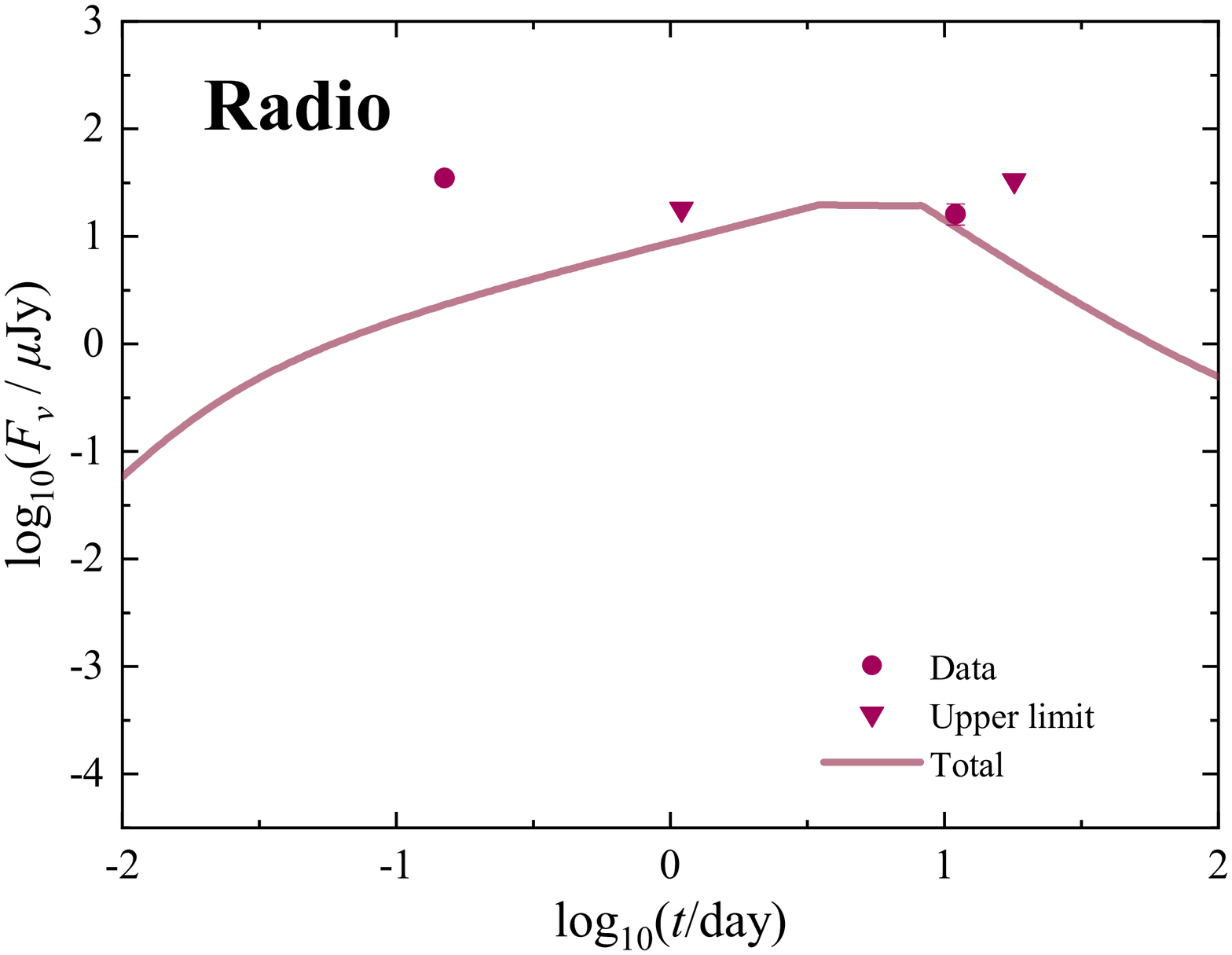}
    \caption{The multi-wavelength light curves of the afterglow emission of GRB\,160821B for different electromagnetic bands as labeled. The observational data are taken from \citep{Lamb2019,Troja2019} and \emph{Swift}. The dotted, dash-dotted, dashed, and solid lines correspond to the emissions from the jet external shock, the thermal kilonova, the PWN, and their combination, respectively. The insert shows the early X-ray afterglow plateau, which may indicate the remnant NS could be intrinsically a magnetar before its surface magnetic field is buried.}
    \label{LC}
\end{figure*}

\begin{figure*}
	\centering
	\includegraphics[width = 0.32\linewidth , trim = 50 30 95 50, clip]{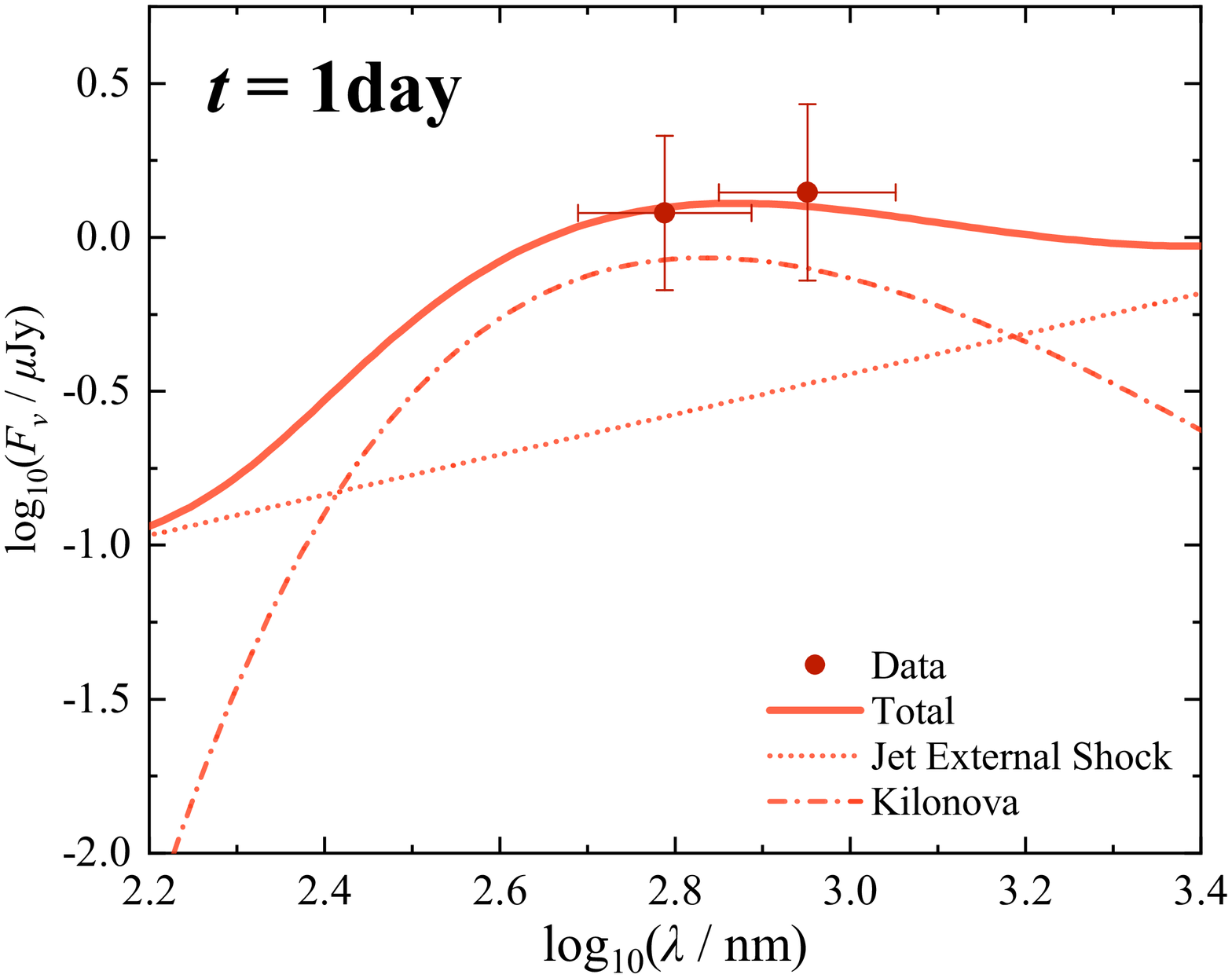}
    \includegraphics[width = 0.32\linewidth , trim = 50 30 95 50, clip]{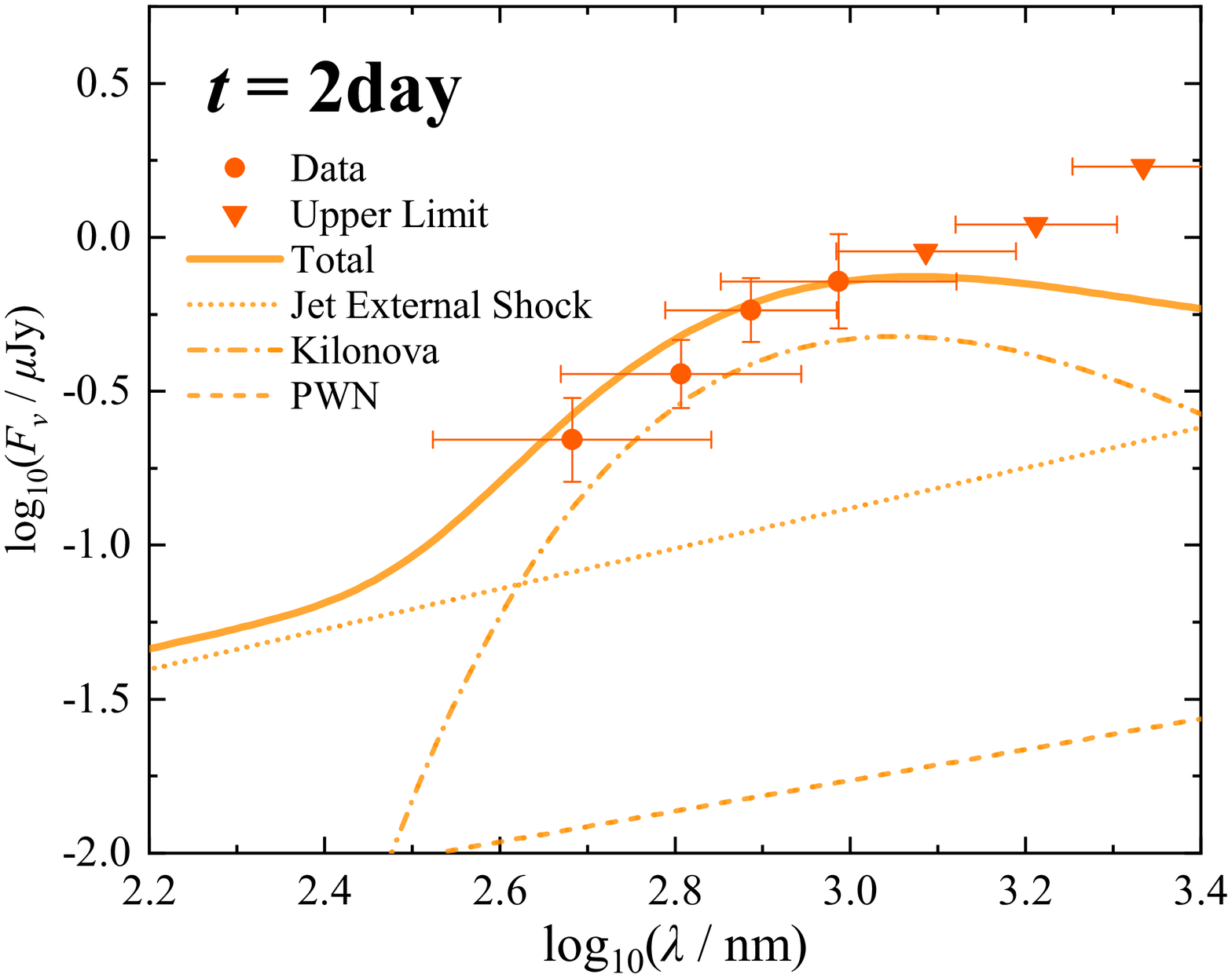}
    \includegraphics[width = 0.32\linewidth , trim = 50 30 95 50, clip]{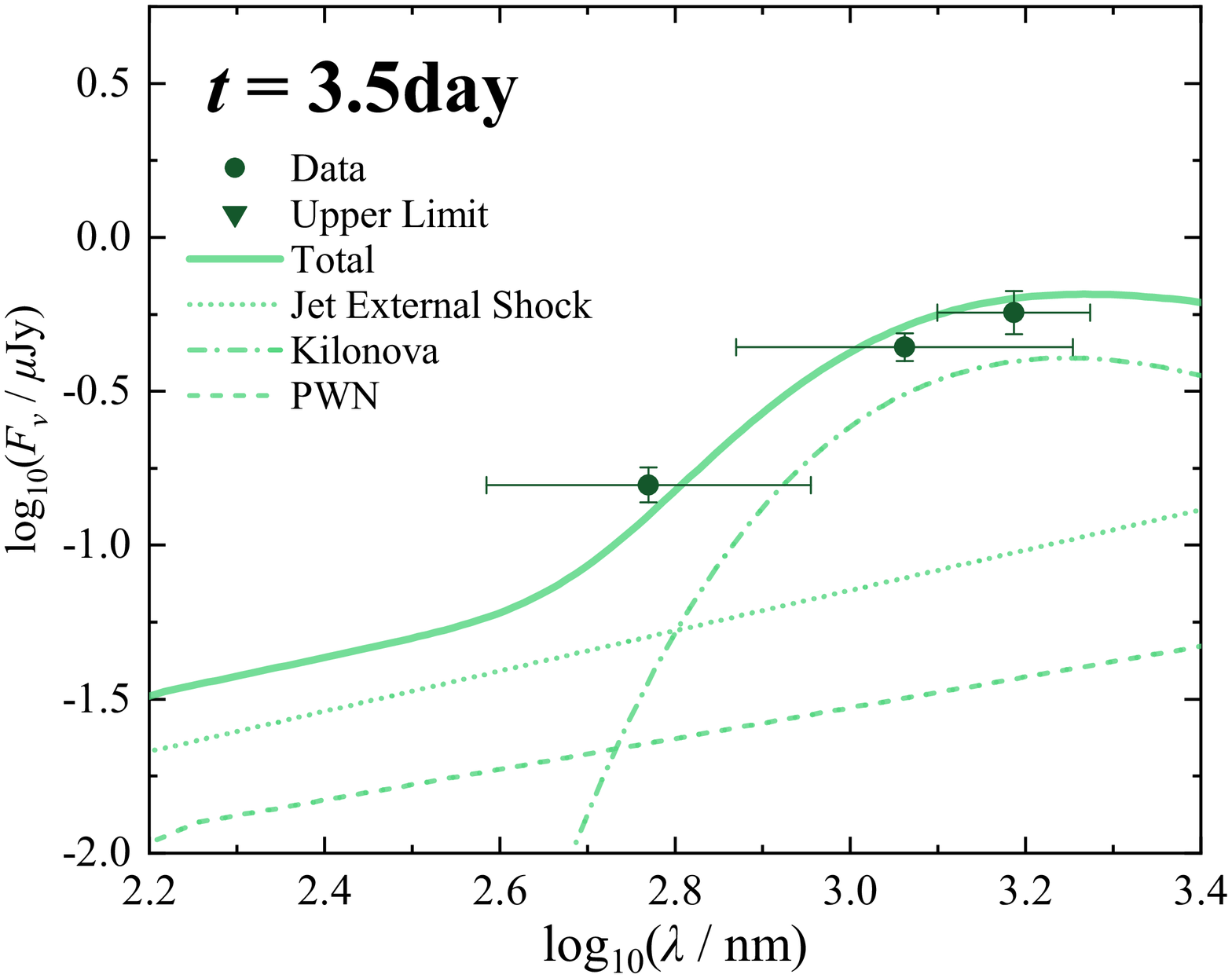}
    \includegraphics[width = 0.32\linewidth , trim = 50 30 95 50, clip]{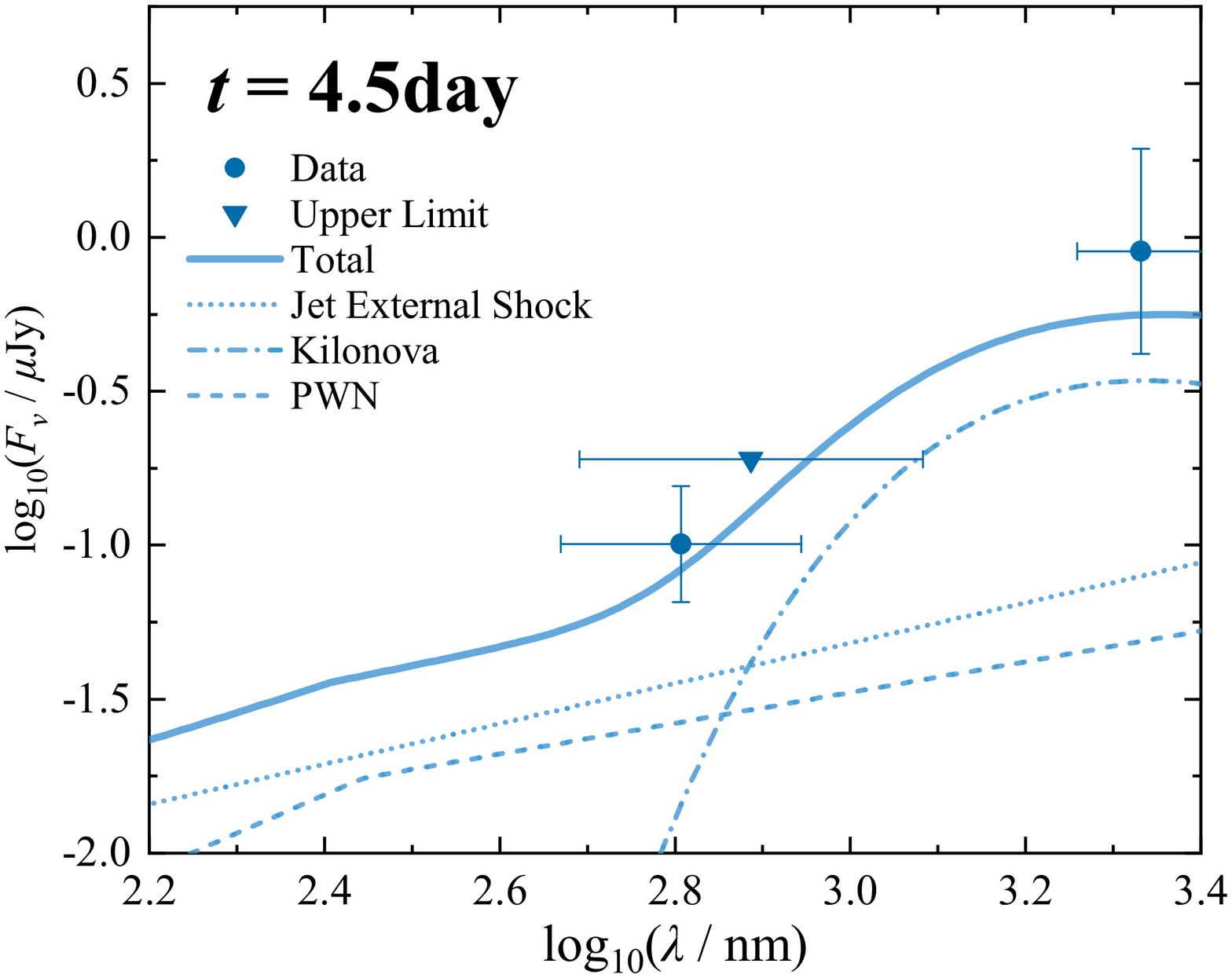}
    \includegraphics[width = 0.32\linewidth , trim = 50 30 95 50, clip]{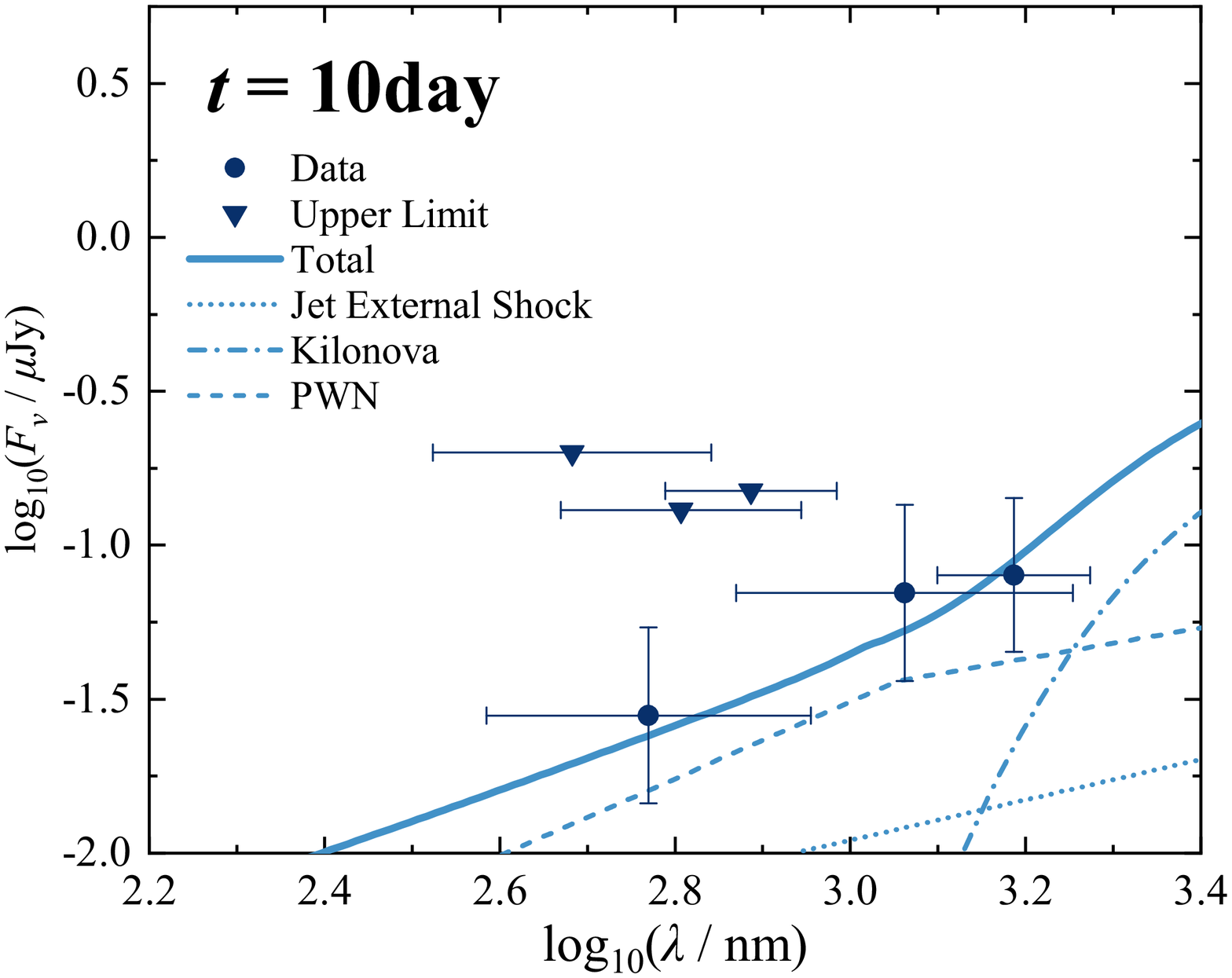}
    \includegraphics[width = 0.32\linewidth , trim = 50 30 95 50, clip]{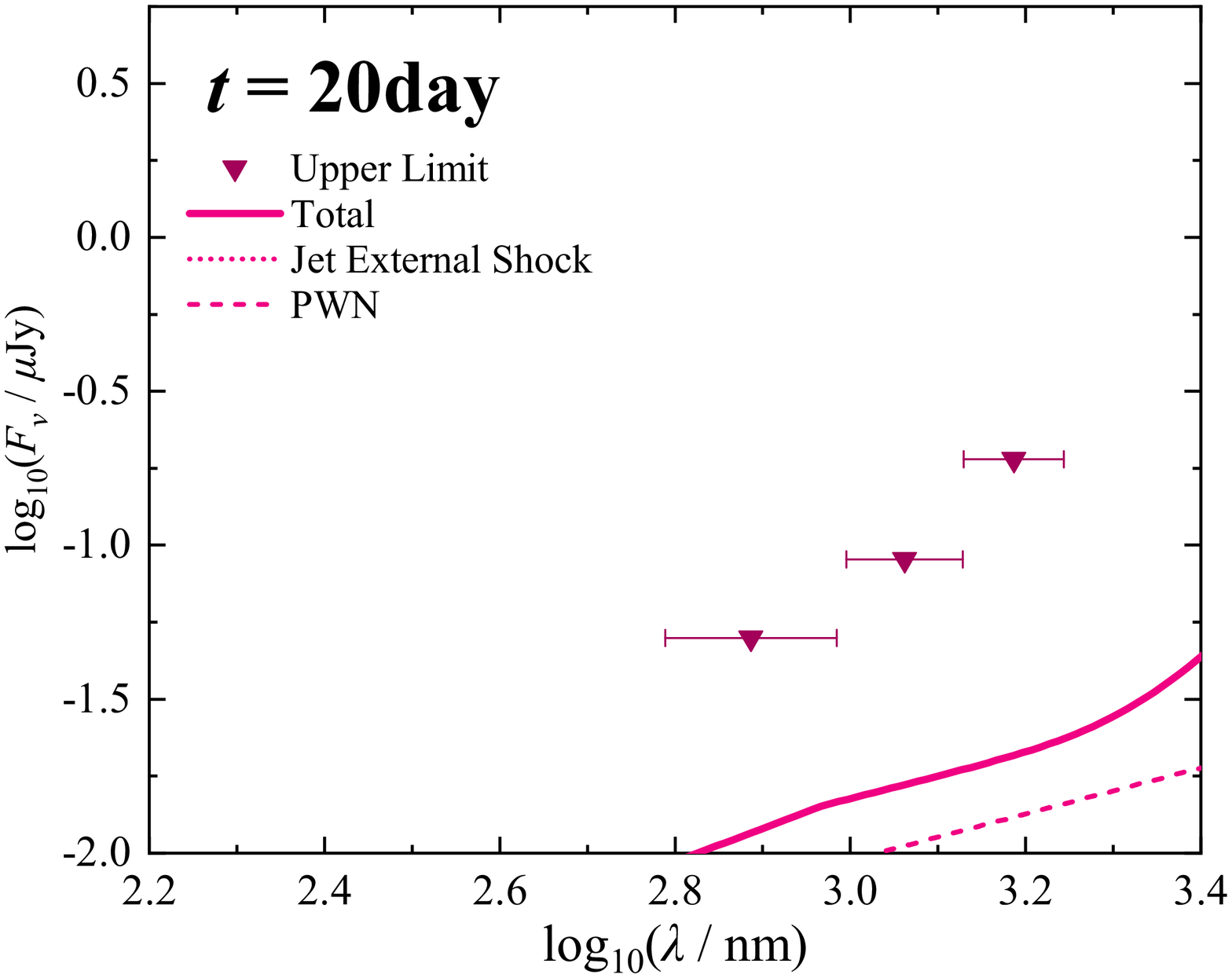}
	\caption{The ultraviolet-optical-infrared spectra of the afterglow emission of GRB\,160821B for different times as labeled. The observational data are taken from \citep{Lamb2019,Troja2019}. The dotted, dash-dotted, dashed, and solid lines correspond to the emissions of the jet external shock, the thermal kilonova, the PWN, and their combination, respectively.}
	\label{Spec}
\end{figure*}

\subsection{The jet ES emission}
The multi-wavelength afterglows of GRB\,160821B must be primarily contributed by the ES driven by the GRB jet decelerating into the interstellar medium. In principle, the relativistic wind from the remnant NS can also interact with the GRB jet and thus influence the emission of the ES, which can be described by the model of \cite{Dai2004} and \cite{Yu2007}. Nevertheless, by considering that the wind luminosity is probably only comparable to the luminosity of the kilonova, the influence of the NS wind on the jet ES can be basically neglected in the case of GRB\,160821B, unless the wind energy is extremely highly collimated in the jet direction (i.e., $\xi\ll1$). Therefore, we simply employ the standard ES model introduced in \cite{Sari1998} and \cite{Huang2000,Huang2003}. In short, while the dynamical evolution of the jet is described by the classical formula as
\begin{equation}\label{}
\Gamma_{\rm j}(t)=\left(17E_{\rm j}\over 1024\pi n_0 m_{\rm p}c^5t^3\right)^{1/8},
\end{equation}
the synchrotron luminosity $L_{\nu}^{\rm jet}$ of the ES emission can also be determined by using Eq. (\ref{Lpwn}) for corresponding parameters and characteristic quantities. Here, $\Gamma_{\rm j}$ and $E_{\rm j}$ are the Lorentz factor and the isotropically-equivalent kinetic energy of the jet, $n_0$ is the particle number density of the medium, and $m_{\rm p}$ is the mass of proton. A half-opening angle $\theta_{\rm j}$ of the jet is taken in our calculation. When the Lorentz factor of the jet decreases to be smaller than $\theta_{\rm j}^{-1}$, the analytical results given by Eq. (\ref{Lpwn}) should be further multiplied by an extra suppression factor $\left(\Gamma_{\rm j}\theta_{\rm j}\right)^{-2}$, which is called as a jet break. Sometimes, the Newtonian effect should also be taken into account.



\section{Comparison with the GRB\,160821B observations}

As described, the electromagnetic emission after a DNS merger is contributed by three components as
\begin{equation}
	L_\nu^{\rm{tot}}=L_\nu^{\rm{jet}}+L_\nu^{\rm kn} + L_ {\nu}^{\rm{pwn}}e^{-\tau_{\nu,\rm tot}},
\end{equation}
which can be used to fit the multi-wavelength afterglow emission of GRB\,160821B. In view of the limited number of the afterglow data, in this paper we do not seek for a completed constraint on the model parameters. Alternatively, we just try to test the compatibility of the model with the observations. A tentative modeling of the light curves and spectra are presented in Figures \ref{LC} and \ref{Spec}, respectively. The used parameter values are listed in Table 1, which are all typical for SGRBs including GRB\,170817A. To be specific, firstly, the X-ray afterglow for $t>10^3$ s can be easily accounted for by the emission from the jet ES, which was not influenced by the energy release from the remnant NS because of the huge kinetic energy of the jet. Moreover, a jet break could happen at a few days after GRB\,160821B. It should also been mentioned that the very early plateau in the X-ray afterglow has not been modeled. This plateau can probably be attributed to the PWN emission at the jet direction before the NS's surface magnetic field was buried (see \cite{Yu2018} and \cite{Li2021} for related discussions). Secondly, for the radio afterglow, the upper limit appearing at $t\sim1$ day indicated that the two observational data could not have the same origin, as previously suggested by \cite{Troja2019} and \cite{Lamb2019}. Then, while the late data at $\sim 10$ day is ascribed to the ES emission, the early data at $\sim 0.1$ day is probably contributed by another emission region, e.g., a reverse shock propagating into the GRB jet (Troja et al. 2019). Finally, for the optical/nIR data, by subtracting the contribution from the jet ES, we can naturally attribute the excesses to the kilonova and PWN emission.

In more detail, the peak time of the kilonova emission is found to be around $\sim1$ day, which can be naturally explained by the diffusion timescale of $ t_{\rm{p}}=\left ( \kappa_{\rm es} M_{\rm{ej}}/4 \pi v_{\rm{ej}} c \right)^{1/2} $ for a typical ejecta mass of $M_{\rm ej}\sim 0.01M_{\odot}$ and a typical electron-scattering opacity of $\kappa_{\rm es} \sim 0.2\, \rm{cm^{2}\,g^{-1}}$. Although a large number of lanthanides are believed to be synthesized in the merger ejecta, the required small opacity can still be understood, by considering that the merger ejecta could be deeply ionized by the early PWN emission before the surface magnetic field of the NS is suppressed. The radioactive power corresponding to the ejecta mass $M_{\rm ej}\sim 0.01M_{\odot}$ can be comparable to the emission luminosity of the PWN. This indicates the kilonova emission after GRB\,160821B was powered by hybrid energy sources, which is the same to the situation of AT\,2017gfo \citep{Yu2018}. In any case, the crucial effect of the remnant NS is mainly displayed in the appearance of the non-thermal PWN emission, which can help to explain the relatively slow decay of the excess emission. Such a non-thermal PWN emission can also be clearly seen from the evolving spectra presented in Figure \ref{Spec}. As shown, while the thermal component shifts from the UV to the IR quickly, the synchrotron emission from the PWN gradually increased to be dominated about $~10\,$days after GRB\,160821B.

\section{Summary}
It is of fundamental importance to determine the nature of the merger product of DNSs. The existence of a long-lived remnant NS can leave many imprints in the kilonova emission after the GW event, by reducing the opacity of the merger ejecta, providing energy injection for the thermal kilonova emission, and, in particularly, contributing a non-thermal emission by the PWN as a result of the interaction between the pulsar wind and the merger ejecta.
Such a situation has been previously suggested to appear in the GRB\,170817A/AT\,2017gfo event. And in this paper, by resolving the non-thermal PWN emission component from the kilonova data, we further find that the remnant NS scenario can also be consistent with the afterglow emission of GRB\,160821B. Although a stringent constraint on the model is still difficult because of the very limited number of the afterglow data, our result still indicates that the formation of a long-lived massive NS is probably not rare in the DNS merger events, which implies the equation of state of the post-merger NSs should be stiff enough. 

\begin{acknowledgements}  This work is supported by the National SKA Program of China  (grant No. 2020SKA0120300) and the National Natural Science Foundation of China  (grant Nos. 11822302 and 11833003)
\end{acknowledgements}

\bibliographystyle{aa}
\bibliography{bib11}

\begin{thebibliography}{92}
\expandafter\ifx\csname natexlab\endcsname\relax\def\natexlab#1{#1}\fi

\bibitem[{Abbott {et~al.}(2017{\natexlab{a}})}]{Abbott2017a}
Abbott, B.~P. {et~al.} 2017{\natexlab{a}}, Astrophys. J. Lett., 848, L13

\bibitem[{Abbott {et~al.}(2017{\natexlab{b}})}]{Abbott2017b}
Abbott, B.~P. {et~al.} 2017{\natexlab{b}}, Phys. Rev. Lett., 119, 161101

\bibitem[{Abbott {et~al.}(2017{\natexlab{c}})}]{Abbott2017c}
Abbott, B.~P. {et~al.} 2017{\natexlab{c}}, Astrophys. J. Lett., 848, L12

\bibitem[{{Alexander} {et~al.}(2017){Alexander}, {Berger}, {Fong}, {Williams},
  {Guidorzi}, {Margutti}, {Metzger}, {Annis}, {Blanchard}, {Brout}, {Brown},
  {Chen}, {Chornock}, {Cowperthwaite}, {Drout}, {Eftekhari}, {Frieman}, {Holz},
  {Nicholl}, {Rest}, {Sako}, {Soares-Santos}, \& {Villar}}]{Alexander2017}
{Alexander}, K.~D., {Berger}, E., {Fong}, W., {et~al.} 2017, \apjl, 848, L21

\bibitem[{{Andreoni} {et~al.}(2017){Andreoni}, {Ackley}, {Cooke}, {Acharyya},
  {Allison}, {Anderson}, {Ashley}, {Baade}, {Bailes}, {Bannister}, {Beardsley},
  {Bessell}, {Bian}, {Bland}, {Boer}, {Booler}, {Brandeker}, {Brown},
  {Buckley}, {Chang}, {Coward}, {Crawford}, {Crisp}, {Crosse}, {Cucchiara},
  {Cup{\'a}k}, {de Gois}, {Deller}, {Devillepoix}, {Dobie}, {Elmer}, {Emrich},
  {Farah}, {Farrell}, {Franzen}, {Gaensler}, {Galloway}, {Gendre}, {Giblin},
  {Goobar}, {Green}, {Hancock}, {Hartig}, {Howell}, {Horsley}, {Hotan},
  {Howie}, {Hu}, {Hu}, {James}, {Johnston}, {Johnston-Hollitt}, {Kaplan},
  {Kasliwal}, {Keane}, {Kenney}, {Klotz}, {Lau}, {Laugier}, {Lenc}, {Li},
  {Liang}, {Lidman}, {Luvaul}, {Lynch}, {Ma}, {Macpherson}, {Mao},
  {McClelland}, {McCully}, {M{\"o}ller}, {Morales}, {Morris}, {Murphy},
  {Noysena}, {Onken}, {Orange}, {Os{\l}owski}, {Pallot}, {Paxman}, {Potter},
  {Pritchard}, {Raja}, {Ridden-Harper}, {Romero-Colmenero}, {Sadler}, {Sansom},
  {Scalzo}, {Schmidt}, {Scott}, {Seghouani}, {Shang}, {Shannon}, {Shao},
  {Shara}, {Sharp}, {Sokolowski}, {Sollerman}, {Staff}, {Steele}, {Sun},
  {Suntzeff}, {Tao}, {Tingay}, {Towner}, {Thierry}, {Trott}, {Tucker},
  {V{\"a}is{\"a}nen}, {Krishnan}, {Walker}, {Wang}, {Wang}, {Wayth}, {Whiting},
  {Williams}, {Williams}, {Wolf}, {Wu}, {Wu}, {Yang}, {Yuan}, {Zhang}, {Zhou},
  \& {Zovaro}}]{Andreoni2017}
{Andreoni}, I., {Ackley}, K., {Cooke}, J., {et~al.} 2017, \pasa, 34, e069

\bibitem[{{Arcavi} {et~al.}(2017){Arcavi}, {Hosseinzadeh}, {Howell}, {McCully},
  {Poznanski}, {Kasen}, {Barnes}, {Zaltzman}, {Vasylyev}, {Maoz}, \&
  {Valenti}}]{Arcavi2017}
{Arcavi}, I., {Hosseinzadeh}, G., {Howell}, D.~A., {et~al.} 2017, \nat, 551, 64

\bibitem[{{Barnes} \& {Kasen}(2013)}]{Barnes2013}
{Barnes}, J. \& {Kasen}, D. 2013, \apj, 775, 18

\bibitem[{{Barnes} {et~al.}(2016){Barnes}, {Kasen}, {Wu}, \&
  {Mart{\'\i}nez-Pinedo}}]{Barnes2016}
{Barnes}, J., {Kasen}, D., {Wu}, M.-R., \& {Mart{\'\i}nez-Pinedo}, G. 2016,
  \apj, 829, 110

\bibitem[{{Berger} {et~al.}(2013){Berger}, {Fong}, \& {Chornock}}]{Berger2013}
{Berger}, E., {Fong}, W., \& {Chornock}, R. 2013, \apjl, 774, L23

\bibitem[{{Chevalier} \& {Li}(2000)}]{Chevalier2000}
{Chevalier}, R.~A. \& {Li}, Z.-Y. 2000, \apj, 536, 195

\bibitem[{{Chornock} {et~al.}(2017){Chornock}, {Berger}, {Kasen},
  {Cowperthwaite}, {Nicholl}, {Villar}, {Alexand er}, {Blanchard}, {Eftekhari},
  {Fong}, {Margutti}, {Williams}, {Annis}, {Brout}, {Brown}, {Chen}, {Drout},
  {Farr}, {Foley}, {Frieman}, {Fryer}, {Herner}, {Holz}, {Kessler}, {Matheson},
  {Metzger}, {Quataert}, {Rest}, {Sako}, {Scolnic}, {Smith}, \&
  {Soares-Santos}}]{Chornock2017}
{Chornock}, R., {Berger}, E., {Kasen}, D., {et~al.} 2017, \apjl, 848, L19

\bibitem[{{Coulter} {et~al.}(2017){Coulter}, {Foley}, {Kilpatrick}, {Drout},
  {Piro}, {Shappee}, {Siebert}, {Simon}, {Ulloa}, {Kasen}, {Madore},
  {Murguia-Berthier}, {Pan}, {Prochaska}, {Ramirez-Ruiz}, {Rest}, \&
  {Rojas-Bravo}}]{Coulter2017}
{Coulter}, D.~A., {Foley}, R.~J., {Kilpatrick}, C.~D., {et~al.} 2017, Science,
  358, 1556

\bibitem[{{Covino} {et~al.}(2017){Covino}, {Wiersema}, {Fan}, {Toma},
  {Higgins}, {Melandri}, {D'Avanzo}, {Mundell}, {Palazzi}, {Tanvir},
  {Bernardini}, {Branchesi}, {Brocato}, {Campana}, {di Serego Alighieri},
  {G{\"o}tz}, {Fynbo}, {Gao}, {Gomboc}, {Gompertz}, {Greiner}, {Hjorth}, {Jin},
  {Kaper}, {Klose}, {Kobayashi}, {Kopac}, {Kouveliotou}, {Levan}, {Mao},
  {Malesani}, {Pian}, {Rossi}, {Salvaterra}, {Starling}, {Steele},
  {Tagliaferri}, {Troja}, {van der Horst}, \& {Wijers}}]{Covino2017}
{Covino}, S., {Wiersema}, K., {Fan}, Y.~Z., {et~al.} 2017, Nature Astronomy, 1,
  791

\bibitem[{{Cowperthwaite} {et~al.}(2017){Cowperthwaite}, {Berger}, {Villar},
  {Metzger}, {Nicholl}, {Chornock}, {Blanchard}, {Fong}, {Margutti},
  {Soares-Santos}, {Alexander}, {Allam}, {Annis}, {Brout}, {Brown}, {Butler},
  {Chen}, {Diehl}, {Doctor}, {Drout}, {Eftekhari}, {Farr}, {Finley}, {Foley},
  {Frieman}, {Fryer}, {Garc{\'\i}a-Bellido}, {Gill}, {Guillochon}, {Herner},
  {Holz}, {Kasen}, {Kessler}, {Marriner}, {Matheson}, {Neilsen}, {Quataert},
  {Palmese}, {Rest}, {Sako}, {Scolnic}, {Smith}, {Tucker}, {Williams},
  {Balbinot}, {Carlin}, {Cook}, {Durret}, {Li}, {Lopes}, {Louren{\c{c}}o},
  {Marshall}, {Medina}, {Muir}, {Mu{\~n}oz}, {Sauseda}, {Schlegel}, {Secco},
  {Vivas}, {Wester}, {Zenteno}, {Zhang}, {Abbott}, {Banerji}, {Bechtol},
  {Benoit-L{\'e}vy}, {Bertin}, {Buckley-Geer}, {Burke}, {Capozzi}, {Carnero
  Rosell}, {Carrasco Kind}, {Castander}, {Crocce}, {Cunha}, {D'Andrea}, {da
  Costa}, {Davis}, {DePoy}, {Desai}, {Dietrich}, {Drlica-Wagner}, {Eifler},
  {Evrard}, {Fernand ez}, {Flaugher}, {Fosalba}, {Gaztanaga}, {Gerdes},
  {Giannantonio}, {Goldstein}, {Gruen}, {Gruendl}, {Gutierrez}, {Honscheid},
  {Jain}, {James}, {Jeltema}, {Johnson}, {Johnson}, {Kent}, {Krause}, {Kron},
  {Kuehn}, {Nuropatkin}, {Lahav}, {Lima}, {Lin}, {Maia}, {March}, {Martini},
  {McMahon}, {Menanteau}, {Miller}, {Miquel}, {Mohr}, {Neilsen}, {Nichol},
  {Ogando}, {Plazas}, {Roe}, {Romer}, {Roodman}, {Rykoff}, {Sanchez},
  {Scarpine}, {Schindler}, {Schubnell}, {Sevilla-Noarbe}, {Smith}, {Smith},
  {Sobreira}, {Suchyta}, {Swanson}, {Tarle}, {Thomas}, {Thomas}, {Troxel},
  {Vikram}, {Walker}, {Wechsler}, {Weller}, {Yanny}, \&
  {Zuntz}}]{Cowperthwaite2017}
{Cowperthwaite}, P.~S., {Berger}, E., {Villar}, V.~A., {et~al.} 2017, \apjl,
  848, L17

\bibitem[{{Dai}(2004)}]{Dai2004}
{Dai}, Z.~G. 2004, \apj, 606, 1000

\bibitem[{{D'Avanzo} {et~al.}(2018){D'Avanzo}, {Campana}, {Salafia}, {Ghirland
  a}, {Ghisellini}, {Melandri}, {Bernardini}, {Branchesi}, {Chassande-Mottin},
  {Covino}, {D'Elia}, {Nava}, {Salvaterra}, {Tagliaferri}, \&
  {Vergani}}]{D'Avanzo2018}
{D'Avanzo}, P., {Campana}, S., {Salafia}, O.~S., {et~al.} 2018, \aap, 613, L1

\bibitem[{{Eichler} {et~al.}(1989){Eichler}, {Livio}, {Piran}, \&
  {Schramm}}]{Eichler1989}
{Eichler}, D., {Livio}, M., {Piran}, T., \& {Schramm}, D.~N. 1989, \nat, 340,
  126

\bibitem[{{Evans} {et~al.}(2017){Evans}, {Cenko}, {Kennea}, {Emery}, {Kuin},
  {Korobkin}, {Wollaeger}, {Fryer}, {Madsen}, {Harrison}, {Xu}, {Nakar},
  {Hotokezaka}, {Lien}, {Campana}, {Oates}, {Troja}, {Breeveld}, {Marshall},
  {Barthelmy}, {Beardmore}, {Burrows}, {Cusumano}, {D'A{\`\i}}, {D'Avanzo},
  {D'Elia}, {de Pasquale}, {Even}, {Fontes}, {Forster}, {Garcia}, {Giommi},
  {Grefenstette}, {Gronwall}, {Hartmann}, {Heida}, {Hungerford}, {Kasliwal},
  {Krimm}, {Levan}, {Malesani}, {Melandri}, {Miyasaka}, {Nousek}, {O'Brien},
  {Osborne}, {Pagani}, {Page}, {Palmer}, {Perri}, {Pike}, {Racusin}, {Rosswog},
  {Siegel}, {Sakamoto}, {Sbarufatti}, {Tagliaferri}, {Tanvir}, \&
  {Tohuvavohu}}]{Evans2017}
{Evans}, P.~A., {Cenko}, S.~B., {Kennea}, J.~A., {et~al.} 2017, Science, 358,
  1565

\bibitem[{{Gao} {et~al.}(2015){Gao}, {Ding}, {Wu}, {Dai}, \& {Zhang}}]{Gao2015}
{Gao}, H., {Ding}, X., {Wu}, X.-F., {Dai}, Z.-G., \& {Zhang}, B. 2015, \apj,
  807, 163

\bibitem[{{Gao} {et~al.}(2017){Gao}, {Zhang}, {L{\"u}}, \& {Li}}]{Gao2017}
{Gao}, H., {Zhang}, B., {L{\"u}}, H.-J., \& {Li}, Y. 2017, \apj, 837, 50

\bibitem[{{Ghirlanda} {et~al.}(2019){Ghirlanda}, {Salafia}, {Paragi},
  {Giroletti}, {Yang}, {Marcote}, {Blanchard}, {Agudo}, {An}, {Bernardini},
  {Beswick}, {Branchesi}, {Campana}, {Casadio}, {Chassand e-Mottin}, {Colpi},
  {Covino}, {D'Avanzo}, {D'Elia}, {Frey}, {Gawronski}, {Ghisellini}, {Gurvits},
  {Jonker}, {van Langevelde}, {Melandri}, {Moldon}, {Nava}, {Perego},
  {Perez-Torres}, {Reynolds}, {Salvaterra}, {Tagliaferri}, {Venturi},
  {Vergani}, \& {Zhang}}]{Ghirlanda2019}
{Ghirlanda}, G., {Salafia}, O.~S., {Paragi}, Z., {et~al.} 2019, Science, 363,
  968

\bibitem[{{Goldstein} {et~al.}(2017){Goldstein}, {Veres}, {Burns}, {Briggs},
  {Hamburg}, {Kocevski}, {Wilson-Hodge}, {Preece}, {Poolakkil}, {Roberts},
  {Hui}, {Connaughton}, {Racusin}, {von Kienlin}, {Dal Canton}, {Christensen},
  {Littenberg}, {Siellez}, {Blackburn}, {Broida}, {Bissaldi}, {Cleveland},
  {Gibby}, {Giles}, {Kippen}, {McBreen}, {McEnery}, {Meegan}, {Paciesas}, \&
  {Stanbro}}]{Goldstein2017}
{Goldstein}, A., {Veres}, P., {Burns}, E., {et~al.} 2017, \apjl, 848, L14

\bibitem[{{Granot} \& {Sari}(2002)}]{Granot2002}
{Granot}, J. \& {Sari}, R. 2002, \apj, 568, 820

\bibitem[{{Grossman} {et~al.}(2014){Grossman}, {Korobkin}, {Rosswog}, \&
  {Piran}}]{Grossman2014}
{Grossman}, D., {Korobkin}, O., {Rosswog}, S., \& {Piran}, T. 2014, \mnras,
  439, 757

\bibitem[{{Hallinan} {et~al.}(2017){Hallinan}, {Corsi}, {Mooley}, {Hotokezaka},
  {Nakar}, {Kasliwal}, {Kaplan}, {Frail}, {Myers}, {Murphy}, {De}, {Dobie},
  {Allison}, {Bannister}, {Bhalerao}, {Chandra}, {Clarke}, {Giacintucci}, {Ho},
  {Horesh}, {Kassim}, {Kulkarni}, {Lenc}, {Lockman}, {Lynch}, {Nichols},
  {Nissanke}, {Palliyaguru}, {Peters}, {Piran}, {Rana}, {Sadler}, \&
  {Singer}}]{Hallinan2017}
{Hallinan}, G., {Corsi}, A., {Mooley}, K.~P., {et~al.} 2017, Science, 358, 1579

\bibitem[{{Hu} {et~al.}(2017){Hu}, {Wu}, {Andreoni}, {Ashley}, {Cooke}, {Cui},
  {Du}, {Dai}, {Gu}, {Hu}, {Lu}, {Li}, {Li}, {Liang}, {Liu}, {Ma}, {Shang},
  {Sun}, {Suntzeff}, {Tao}, {Udden}, {Wang}, {Wang}, {Wen}, {Xiao}, {Su},
  {Yang}, {Yang}, {Yuan}, {Zhou}, {Zhang}, {Zhou}, \& {Zhu}}]{Hu2017}
{Hu}, L., {Wu}, X., {Andreoni}, I., {et~al.} 2017, Science Bulletin, 62, 1433

\bibitem[{{Huang} \& {Cheng}(2003)}]{Huang2003}
{Huang}, Y.~F. \& {Cheng}, K.~S. 2003, \mnras, 341, 263

\bibitem[{{Huang} {et~al.}(2000){Huang}, {Gou}, {Dai}, \& {Lu}}]{Huang2000}
{Huang}, Y.~F., {Gou}, L.~J., {Dai}, Z.~G., \& {Lu}, T. 2000, \apj, 543, 90

\bibitem[{{Jin} {et~al.}(2020){Jin}, {Covino}, {Liao}, {Li}, {D'Avanzo}, {Fan},
  \& {Wei}}]{Jin2020}
{Jin}, Z.-P., {Covino}, S., {Liao}, N.-H., {et~al.} 2020, Nature Astronomy, 4,
  77

\bibitem[{{Jin} {et~al.}(2016){Jin}, {Hotokezaka}, {Li}, {Tanaka}, {D'Avanzo},
  {Fan}, {Covino}, {Wei}, \& {Piran}}]{Jin2016}
{Jin}, Z.-P., {Hotokezaka}, K., {Li}, X., {et~al.} 2016, Nature Communications,
  7, 12898

\bibitem[{{Jin} {et~al.}(2015){Jin}, {Li}, {Cano}, {Covino}, {Fan}, \&
  {Wei}}]{Jin2015}
{Jin}, Z.-P., {Li}, X., {Cano}, Z., {et~al.} 2015, \apjl, 811, L22

\bibitem[{{Jin} {et~al.}(2018){Jin}, {Li}, {Wang}, {Wang}, {He}, {Yuan},
  {Zhang}, {Zou}, {Fan}, \& {Wei}}]{Jin2018}
{Jin}, Z.-P., {Li}, X., {Wang}, H., {et~al.} 2018, \apj, 857, 128

\bibitem[{{Kasen} {et~al.}(2013){Kasen}, {Badnell}, \& {Barnes}}]{Kasen2013}
{Kasen}, D., {Badnell}, N.~R., \& {Barnes}, J. 2013, \apj, 774, 25

\bibitem[{{Kasen} \& {Bildsten}(2010)}]{Kasen2010}
{Kasen}, D. \& {Bildsten}, L. 2010, \apj, 717, 245

\bibitem[{{Kasen} {et~al.}(2015){Kasen}, {Fern{\'a}ndez}, \&
  {Metzger}}]{Kasen2015}
{Kasen}, D., {Fern{\'a}ndez}, R., \& {Metzger}, B.~D. 2015, \mnras, 450, 1777

\bibitem[{{Kasen} {et~al.}(2017){Kasen}, {Metzger}, {Barnes}, {Quataert}, \&
  {Ramirez-Ruiz}}]{Kasen2017}
{Kasen}, D., {Metzger}, B., {Barnes}, J., {Quataert}, E., \& {Ramirez-Ruiz}, E.
  2017, \nat, 551, 80

\bibitem[{{Kasliwal} {et~al.}(2017){Kasliwal}, {Korobkin}, {Lau}, {Wollaeger},
  \& {Fryer}}]{Kasliwal2017}
{Kasliwal}, M.~M., {Korobkin}, O., {Lau}, R.~M., {Wollaeger}, R., \& {Fryer},
  C.~L. 2017, \apjl, 843, L34

\bibitem[{{Kawaguchi} {et~al.}(2018){Kawaguchi}, {Shibata}, \&
  {Tanaka}}]{Kawaguchi2018}
{Kawaguchi}, K., {Shibata}, M., \& {Tanaka}, M. 2018, \apjl, 865, L21

\bibitem[{{Kilpatrick} {et~al.}(2017){Kilpatrick}, {Foley}, {Kasen},
  {Murguia-Berthier}, {Ramirez-Ruiz}, {Coulter}, {Drout}, {Piro}, {Shappee},
  {Boutsia}, {Contreras}, {Di Mille}, {Madore}, {Morrell}, {Pan}, {Prochaska},
  {Rest}, {Rojas-Bravo}, {Siebert}, {Simon}, \& {Ulloa}}]{Kilpatrick2017}
{Kilpatrick}, C.~D., {Foley}, R.~J., {Kasen}, D., {et~al.} 2017, Science, 358,
  1583

\bibitem[{{Korobkin} {et~al.}(2012){Korobkin}, {Rosswog}, {Arcones}, \&
  {Winteler}}]{Korobkin2012}
{Korobkin}, O., {Rosswog}, S., {Arcones}, A., \& {Winteler}, C. 2012, \mnras,
  426, 1940

\bibitem[{{Kotera} {et~al.}(2013){Kotera}, {Phinney}, \& {Olinto}}]{Kotera2013}
{Kotera}, K., {Phinney}, E.~S., \& {Olinto}, A.~V. 2013, \mnras, 432, 3228

\bibitem[{{Lamb} {et~al.}(2019){Lamb}, {Tanvir}, {Levan}, {de Ugarte Postigo},
  {Kawaguchi}, {Corsi}, {Evans}, {Gompertz}, {Malesani}, {Page}, {Wiersema},
  {Rosswog}, {Shibata}, {Tanaka}, {van der Horst}, {Cano}, {Fynbo}, {Fruchter},
  {Greiner}, {Heintz}, {Higgins}, {Hjorth}, {Izzo}, {Jakobsson}, {Kann},
  {O'Brien}, {Perley}, {Pian}, {Pugliese}, {Starling}, {Th{\"o}ne}, {Watson},
  {Wijers}, \& {Xu}}]{Lamb2019}
{Lamb}, G.~P., {Tanvir}, N.~R., {Levan}, A.~J., {et~al.} 2019, \apj, 883, 48

\bibitem[{{Lattimer} \& {Schramm}(1974)}]{Lattimer1974}
{Lattimer}, J.~M. \& {Schramm}, D.~N. 1974, \apjl, 192, L145

\bibitem[{{Lattimer} \& {Schramm}(1976)}]{Lattimer1976}
{Lattimer}, J.~M. \& {Schramm}, D.~N. 1976, \apj, 210, 549

\bibitem[{{Lazzati} {et~al.}(2018){Lazzati}, {Perna}, {Morsony},
  {Lopez-Camara}, {Cantiello}, {Ciolfi}, {Giacomazzo}, \&
  {Workman}}]{Lazzati2018}
{Lazzati}, D., {Perna}, R., {Morsony}, B.~J., {et~al.} 2018, \prl, 120, 241103

\bibitem[{{Li} \& {Paczy{\'n}ski}(1998)}]{Li1998}
{Li}, L.-X. \& {Paczy{\'n}ski}, B. 1998, \apjl, 507, L59

\bibitem[{{Li} {et~al.}(2018){Li}, {Liu}, {Yu}, \& {Zhang}}]{Li2018}
{Li}, S.-Z., {Liu}, L.-D., {Yu}, Y.-W., \& {Zhang}, B. 2018, \apjl, 861, L12

\bibitem[{{Li} \& {Yu}(2016)}]{Li2016}
{Li}, S.-Z. \& {Yu}, Y.-W. 2016, \apj, 819, 120

\bibitem[{{Li} {et~al.}(2021){Li}, {Yu}, {Gao}, \& {Zhang}}]{li2021}
{Li}, S.-Z., {Yu}, Y.-W., {Gao}, H., \& {Zhang}, B. 2021, \apj, 907, 87

\bibitem[{{Lipunov} {et~al.}(2017){Lipunov}, {Gorbovskoy}, {Kornilov}, {.
  Tyurina}, {Balanutsa}, {Kuznetsov}, {Vlasenko}, {Kuvshinov}, {Gorbunov},
  {Buckley}, {Krylov}, {Podesta}, {Lopez}, {Podesta}, {Levato}, {Saffe},
  {Mallamachi}, {Potter}, {Budnev}, {Gress}, {Ishmuhametova}, {Vladimirov},
  {Zimnukhov}, {Yurkov}, {Sergienko}, {Gabovich}, {Rebolo}, {Serra-Ricart},
  {Israelyan}, {Chazov}, {Wang}, {Tlatov}, \& {Panchenko}}]{Lipunov2017}
{Lipunov}, V.~M., {Gorbovskoy}, E., {Kornilov}, V.~G., {et~al.} 2017, \apjl,
  850, L1

\bibitem[{{Lyman} {et~al.}(2018){Lyman}, {Lamb}, {Levan}, {Mandel}, {Tanvir},
  {Kobayashi}, {Gompertz}, {Hjorth}, {Fruchter}, {Kangas}, {Steeghs}, {Steele},
  {Cano}, {Copperwheat}, {Evans}, {Fynbo}, {Gall}, {Im}, {Izzo}, {Jakobsson},
  {Milvang-Jensen}, {O'Brien}, {Osborne}, {Palazzi}, {Perley}, {Pian},
  {Rosswog}, {Rowlinson}, {Schulze}, {Stanway}, {Sutton}, {Th{\"o}ne}, {de
  Ugarte Postigo}, {Watson}, {Wiersema}, \& {Wijers}}]{Lyman2018}
{Lyman}, J.~D., {Lamb}, G.~P., {Levan}, A.~J., {et~al.} 2018, Nature Astronomy,
  2, 751

\bibitem[{{Ma} {et~al.}(2020){Ma}, {Xie}, {Liao}, {Zhang}, {L{\"u}}, {Liu}, \&
  {Lei}}]{Ma2020}
{Ma}, S.-B., {Xie}, W., {Liao}, B., {et~al.} 2020, arXiv e-prints,
  arXiv:2010.01338

\bibitem[{{Margutti} {et~al.}(2017){Margutti}, {Berger}, {Fong}, {Guidorzi},
  {Alexander}, {Metzger}, {Blanchard}, {Cowperthwaite}, {Chornock},
  {Eftekhari}, {Nicholl}, {Villar}, {Williams}, {Annis}, {Brown}, {Chen},
  {Doctor}, {Frieman}, {Holz}, {Sako}, \& {Soares-Santos}}]{Margutti2017}
{Margutti}, R., {Berger}, E., {Fong}, W., {et~al.} 2017, \apjl, 848, L20

\bibitem[{{Martin} {et~al.}(2015){Martin}, {Perego}, {Arcones}, {Thielemann},
  {Korobkin}, \& {Rosswog}}]{Martin2015}
{Martin}, D., {Perego}, A., {Arcones}, A., {et~al.} 2015, \apj, 813, 2

\bibitem[{{M{\'e}sz{\'a}ros} \& {Rees}(1997)}]{Mszros1997}
{M{\'e}sz{\'a}ros}, P. \& {Rees}, M.~J. 1997, \apj, 476, 232

\bibitem[{{Metzger}(2017{\natexlab{a}})}]{Metzger2017}
{Metzger}, B.~D. 2017{\natexlab{a}}, Living Reviews in Relativity, 20, 3

\bibitem[{{Metzger}(2017{\natexlab{b}})}]{Metzger2017arXiv}
{Metzger}, B.~D. 2017{\natexlab{b}}, arXiv e-prints, arXiv:1710.05931

\bibitem[{{Metzger} \& {Fern{\'a}ndez}(2014)}]{Metzger2014b}
{Metzger}, B.~D. \& {Fern{\'a}ndez}, R. 2014, \mnras, 441, 3444

\bibitem[{{Metzger} {et~al.}(2010){Metzger}, {Mart{\'\i}nez-Pinedo}, {Darbha},
  {Quataert}, {Arcones}, {Kasen}, {Thomas}, {Nugent}, {Panov}, \&
  {Zinner}}]{Metzger2010}
{Metzger}, B.~D., {Mart{\'\i}nez-Pinedo}, G., {Darbha}, S., {et~al.} 2010,
  \mnras, 406, 2650

\bibitem[{{Metzger} \& {Piro}(2014)}]{Metzger2014a}
{Metzger}, B.~D. \& {Piro}, A.~L. 2014, \mnras, 439, 3916

\bibitem[{{Nagakura} {et~al.}(2014){Nagakura}, {Hotokezaka}, {Sekiguchi},
  {Shibata}, \& {Ioka}}]{Nagakura2014}
{Nagakura}, H., {Hotokezaka}, K., {Sekiguchi}, Y., {Shibata}, M., \& {Ioka}, K.
  2014, \apjl, 784, L28

\bibitem[{{Nakar}(2007)}]{Nakar2007}
{Nakar}, E. 2007, \physrep, 442, 166

\bibitem[{{Narayan} {et~al.}(1992){Narayan}, {Paczynski}, \&
  {Piran}}]{Narayan1992}
{Narayan}, R., {Paczynski}, B., \& {Piran}, T. 1992, \apjl, 395, L83

\bibitem[{{Nicholl} {et~al.}(2017){Nicholl}, {Berger}, {Kasen}, {Metzger},
  {Elias}, {Brice{\~n}o}, {Alexander}, {Blanchard}, {Chornock},
  {Cowperthwaite}, {Eftekhari}, {Fong}, {Margutti}, {Villar}, {Williams},
  {Brown}, {Annis}, {Bahramian}, {Brout}, {Brown}, {Chen}, {Clemens},
  {Dennihy}, {Dunlap}, {Holz}, {Marchesini}, {Massaro}, {Moskowitz},
  {Pelisoli}, {Rest}, {Ricci}, {Sako}, {Soares-Santos}, \&
  {Strader}}]{Nicholl2017}
{Nicholl}, M., {Berger}, E., {Kasen}, D., {et~al.} 2017, \apjl, 848, L18

\bibitem[{{Paczynski}(1986)}]{Paczynski1986}
{Paczynski}, B. 1986, \apjl, 308, L43

\bibitem[{{Paschalidis} {et~al.}(2015){Paschalidis}, {Ruiz}, \&
  {Shapiro}}]{Paschalidis2015}
{Paschalidis}, V., {Ruiz}, M., \& {Shapiro}, S.~L. 2015, \apjl, 806, L14

\bibitem[{{Perego} {et~al.}(2017){Perego}, {Radice}, \&
  {Bernuzzi}}]{Perego2017}
{Perego}, A., {Radice}, D., \& {Bernuzzi}, S. 2017, \apjl, 850, L37

\bibitem[{{Perego} {et~al.}(2014){Perego}, {Rosswog}, {Cabez{\'o}n},
  {Korobkin}, {K{\"a}ppeli}, {Arcones}, \& {Liebend{\"o}rfer}}]{Perego2014}
{Perego}, A., {Rosswog}, S., {Cabez{\'o}n}, R.~M., {et~al.} 2014, \mnras, 443,
  3134

\bibitem[{{Piro} {et~al.}(2019){Piro}, {Troja}, {Zhang}, {Ryan}, {van Eerten},
  {Ricci}, {Wieringa}, {Tiengo}, {Butler}, {Cenko}, {Fox}, {Khandrika},
  {Novara}, {Rossi}, \& {Sakamoto}}]{Piro2019}
{Piro}, L., {Troja}, E., {Zhang}, B., {et~al.} 2019, \mnras, 483, 1912

\bibitem[{{Rees} \& {Meszaros}(1992)}]{Rees1992}
{Rees}, M.~J. \& {Meszaros}, P. 1992, \mnras, 258, 41

\bibitem[{{Ren} {et~al.}(2019){Ren}, {Lin}, {Zhang}, {Li}, {Liu}, {Lu}, {Wang},
  \& {Liang}}]{Ren2019}
{Ren}, J., {Lin}, D.-B., {Zhang}, L.-L., {et~al.} 2019, \apj, 885, 60

\bibitem[{{Rezzolla} {et~al.}(2011){Rezzolla}, {Giacomazzo}, {Baiotti},
  {Granot}, {Kouveliotou}, \& {Aloy}}]{Rezzolla2011}
{Rezzolla}, L., {Giacomazzo}, B., {Baiotti}, L., {et~al.} 2011, \apjl, 732, L6

\bibitem[{{Roberts} {et~al.}(2011){Roberts}, {Kasen}, {Lee}, \&
  {Ramirez-Ruiz}}]{Roberts2011}
{Roberts}, L.~F., {Kasen}, D., {Lee}, W.~H., \& {Ramirez-Ruiz}, E. 2011, \apjl,
  736, L21

\bibitem[{{Sari} {et~al.}(1998){Sari}, {Piran}, \& {Narayan}}]{Sari1998}
{Sari}, R., {Piran}, T., \& {Narayan}, R. 1998, \apjl, 497, L17

\bibitem[{{Smartt} {et~al.}(2017){Smartt}, {Chen}, {Jerkstrand}, {Coughlin},
  {Kankare}, {Sim}, {Fraser}, {Inserra}, {Maguire}, {Chambers}, {Huber},
  {Kr{\"u}hler}, {Leloudas}, {Magee}, {Shingles}, {Smith}, {Young}, {Tonry},
  {Kotak}, {Gal-Yam}, {Lyman}, {Homan}, {Agliozzo}, {Anderson}, {Angus},
  {Ashall}, {Barbarino}, {Bauer}, {Berton}, {Botticella}, {Bulla}, {Bulger},
  {Cannizzaro}, {Cano}, {Cartier}, {Cikota}, {Clark}, {De Cia}, {Della Valle},
  {Denneau}, {Dennefeld}, {Dessart}, {Dimitriadis}, {Elias-Rosa}, {Firth},
  {Flewelling}, {Fl{\"o}rs}, {Franckowiak}, {Frohmaier}, {Galbany},
  {Gonz{\'a}lez-Gait{\'a}n}, {Greiner}, {Gromadzki}, {Guelbenzu},
  {Guti{\'e}rrez}, {Hamanowicz}, {Hanlon}, {Harmanen}, {Heintz}, {Heinze},
  {Hernandez}, {Hodgkin}, {Hook}, {Izzo}, {James}, {Jonker}, {Kerzendorf},
  {Klose}, {Kostrzewa-Rutkowska}, {Kowalski}, {Kromer}, {Kuncarayakti},
  {Lawrence}, {Lowe}, {Magnier}, {Manulis}, {Martin-Carrillo}, {Mattila},
  {McBrien}, {M{\"u}ller}, {Nordin}, {O'Neill}, {Onori}, {Palmerio},
  {Pastorello}, {Patat}, {Pignata}, {Podsiadlowski}, {Pumo}, {Prentice}, {Rau},
  {Razza}, {Rest}, {Reynolds}, {Roy}, {Ruiter}, {Rybicki}, {Salmon}, {Schady},
  {Schultz}, {Schweyer}, {Seitenzahl}, {Smith}, {Sollerman}, {Stalder},
  {Stubbs}, {Sullivan}, {Szegedi}, {Taddia}, {Taubenberger}, {Terreran}, {van
  Soelen}, {Vos}, {Wainscoat}, {Walton}, {Waters}, {Weiland}, {Willman},
  {Wiseman}, {Wright}, {Wyrzykowski}, \& {Yaron}}]{Smartt2017}
{Smartt}, S.~J., {Chen}, T.~W., {Jerkstrand}, A., {et~al.} 2017, \nat, 551, 75

\bibitem[{{Soares-Santos} {et~al.}(2017){Soares-Santos}, {Holz}, {Annis},
  {Chornock}, {Herner}, {Berger}, {Brout}, {Chen}, {Kessler}, {Sako}, {Allam},
  {Tucker}, {Butler}, {Palmese}, {Doctor}, {Diehl}, {Frieman}, {Yanny}, {Lin},
  {Scolnic}, {Cowperthwaite}, {Neilsen}, {Marriner}, {Kuropatkin}, {Hartley},
  {Paz-Chinch{\'o}n}, {Alexander}, {Balbinot}, {Blanchard}, {Brown}, {Carlin},
  {Conselice}, {Cook}, {Drlica-Wagner}, {Drout}, {Durret}, {Eftekhari}, {Farr},
  {Finley}, {Foley}, {Fong}, {Fryer}, {Garc{\'\i}a-Bellido}, {Gill}, {Gruendl},
  {Hanna}, {Kasen}, {Li}, {Lopes}, {Louren{\c{c}}o}, {Margutti}, {Marshall},
  {Matheson}, {Medina}, {Metzger}, {Mu{\~n}oz}, {Muir}, {Nicholl}, {Quataert},
  {Rest}, {Sauseda}, {Schlegel}, {Secco}, {Sobreira}, {Stebbins}, {Villar},
  {Vivas}, {Walker}, {Wester}, {Williams}, {Zenteno}, {Zhang}, {Abbott},
  {Abdalla}, {Banerji}, {Bechtol}, {Benoit-L{\'e}vy}, {Bertin}, {Brooks},
  {Buckley-Geer}, {Burke}, {Carnero Rosell}, {Carrasco Kind}, {Carretero},
  {Castander}, {Crocce}, {Cunha}, {D'Andrea}, {da Costa}, {Davis}, {Desai},
  {Dietrich}, {Doel}, {Eifler}, {Fernand ez}, {Flaugher}, {Fosalba},
  {Gaztanaga}, {Gerdes}, {Giannantonio}, {Goldstein}, {Gruen}, {Gschwend},
  {Gutierrez}, {Honscheid}, {Jain}, {James}, {Jeltema}, {Johnson}, {Johnson},
  {Kent}, {Krause}, {Kron}, {Kuehn}, {Kuhlmann}, {Lahav}, {Lima}, {Maia},
  {March}, {McMahon}, {Menanteau}, {Miquel}, {Mohr}, {Nichol}, {Nord}, {Ogand
  o}, {Petravick}, {Plazas}, {Romer}, {Roodman}, {Rykoff}, {Sanchez},
  {Scarpine}, {Schubnell}, {Sevilla-Noarbe}, {Smith}, {Smith}, {Suchyta},
  {Swanson}, {Tarle}, {Thomas}, {Thomas}, {Troxel}, {Vikram}, {Wechsler},
  {Weller}, {Dark Energy Survey}, \& {Dark Energy Camera GW-EM
  Collaboration}}]{Soares-Santos2017}
{Soares-Santos}, M., {Holz}, D.~E., {Annis}, J., {et~al.} 2017, \apjl, 848, L16

\bibitem[{{Symbalisty} \& {Schramm}(1982)}]{Symbalisty1982}
{Symbalisty}, E. \& {Schramm}, D.~N. 1982, \aplett, 22, 143

\bibitem[{{Tanaka} \& {Hotokezaka}(2013)}]{Tanaka2013}
{Tanaka}, M. \& {Hotokezaka}, K. 2013, \apj, 775, 113

\bibitem[{{Tanaka} {et~al.}(2017){Tanaka}, {Utsumi}, {Mazzali}, {Tominaga},
  {Yoshida}, {Sekiguchi}, {Morokuma}, {Motohara}, {Ohta}, {Kawabata}, {Abe},
  {Aoki}, {Asakura}, {Baar}, {Barway}, {Bond}, {Doi}, {Fujiyoshi}, {Furusawa},
  {Honda}, {Itoh}, {Kawabata}, {Kawai}, {Kim}, {Lee}, {Miyazaki}, {Morihana},
  {Nagashima}, {Nagayama}, {Nakaoka}, {Nakata}, {Ohsawa}, {Ohshima}, {Okita},
  {Saito}, {Sumi}, {Tajitsu}, {Takahashi}, {Takayama}, {Tamura}, {Tanaka},
  {Terai}, {Tristram}, {Yasuda}, \& {Zenko}}]{Tanaka2017}
{Tanaka}, M., {Utsumi}, Y., {Mazzali}, P.~A., {et~al.} 2017, \pasj, 69, 102

\bibitem[{{Tanvir} {et~al.}(2013){Tanvir}, {Levan}, {Fruchter}, {Hjorth},
  {Hounsell}, {Wiersema}, \& {Tunnicliffe}}]{Tanvir2013}
{Tanvir}, N.~R., {Levan}, A.~J., {Fruchter}, A.~S., {et~al.} 2013, \nat, 500,
  547

\bibitem[{{Tanvir} {et~al.}(2017){Tanvir}, {Levan},
  {Gonz{\'a}lez-Fern{\'a}ndez}, {Korobkin}, {Mandel}, {Rosswog}, {Hjorth},
  {D'Avanzo}, {Fruchter}, {Fryer}, {Kangas}, {Milvang-Jensen}, {Rosetti},
  {Steeghs}, {Wollaeger}, {Cano}, {Copperwheat}, {Covino}, {D'Elia}, {de Ugarte
  Postigo}, {Evans}, {Even}, {Fairhurst}, {Figuera Jaimes}, {Fontes}, {Fujii},
  {Fynbo}, {Gompertz}, {Greiner}, {Hodosan}, {Irwin}, {Jakobsson},
  {J{\o}rgensen}, {Kann}, {Lyman}, {Malesani}, {McMahon}, {Melandri},
  {O'Brien}, {Osborne}, {Palazzi}, {Perley}, {Pian}, {Piranomonte}, {Rabus},
  {Rol}, {Rowlinson}, {Schulze}, {Sutton}, {Th{\"o}ne}, {Ulaczyk}, {Watson},
  {Wiersema}, \& {Wijers}}]{Tanvir2017}
{Tanvir}, N.~R., {Levan}, A.~J., {Gonz{\'a}lez-Fern{\'a}ndez}, C., {et~al.}
  2017, \apjl, 848, L27

\bibitem[{{Troja} {et~al.}(2019){Troja}, {Castro-Tirado}, {Becerra
  Gonz{\'a}lez}, {Hu}, {Ryan}, {Cenko}, {Ricci}, {Novara},
  {S{\'a}nchez-R{\'a}mirez}, {Acosta-Pulido}, {Ackley}, {Caballero
  Garc{\'\i}a}, {Eikenberry}, {Guziy}, {Jeong}, {Lien}, {M{\'a}rquez}, {Pand
  ey}, {Park}, {Sakamoto}, {Tello}, {Sokolov}, {Sokolov}, {Tiengo}, {Valeev},
  {Zhang}, \& {Veilleux}}]{Troja2019}
{Troja}, E., {Castro-Tirado}, A.~J., {Becerra Gonz{\'a}lez}, J., {et~al.} 2019,
  \mnras, 489, 2104

\bibitem[{{Troja} {et~al.}(2017){Troja}, {Piro}, {van Eerten}, {Wollaeger},
  {Im}, {Fox}, {Butler}, {Cenko}, {Sakamoto}, {Fryer}, {Ricci}, {Lien}, {Ryan},
  {Korobkin}, {Lee}, {Burgess}, {Lee}, {Watson}, {Choi}, {Covino}, {D'Avanzo},
  {Fontes}, {Gonz{\'a}lez}, {Khandrika}, {Kim}, {Kim}, {Lee}, {Lee}, {Kutyrev},
  {Lim}, {S{\'a}nchez-Ram{\'\i}rez}, {Veilleux}, {Wieringa}, \&
  {Yoon}}]{Troja2017}
{Troja}, E., {Piro}, L., {van Eerten}, H., {et~al.} 2017, \nat, 551, 71

\bibitem[{{Villar} {et~al.}(2017){Villar}, {Guillochon}, {Berger}, {Metzger},
  {Cowperthwaite}, {Nicholl}, {Alexand er}, {Blanchard}, {Chornock},
  {Eftekhari}, {Fong}, {Margutti}, \& {Williams}}]{Villar2017}
{Villar}, V.~A., {Guillochon}, J., {Berger}, E., {et~al.} 2017, \apjl, 851, L21

\bibitem[{{Wanajo} {et~al.}(2014){Wanajo}, {Sekiguchi}, {Nishimura}, {Kiuchi},
  {Kyutoku}, \& {Shibata}}]{Wanajo2014}
{Wanajo}, S., {Sekiguchi}, Y., {Nishimura}, N., {et~al.} 2014, \apjl, 789, L39

\bibitem[{{Yang} {et~al.}(2015){Yang}, {Jin}, {Li}, {Covino}, {Zheng},
  {Hotokezaka}, {Fan}, {Piran}, \& {Wei}}]{Yang2015}
{Yang}, B., {Jin}, Z.-P., {Li}, X., {et~al.} 2015, Nature Communications, 6,
  7323

\bibitem[{{Yu} {et~al.}(2019){Yu}, {Chen}, \& {Li}}]{Yu2019b}
{Yu}, Y.-W., {Chen}, A., \& {Li}, X.-D. 2019, \apjl, 877, L21

\bibitem[{{Yu} \& {Dai}(2007)}]{Yu2007}
{Yu}, Y.~W. \& {Dai}, Z.~G. 2007, \aap, 470, 119

\bibitem[{{Yu} {et~al.}(2018){Yu}, {Liu}, \& {Dai}}]{Yu2018}
{Yu}, Y.-W., {Liu}, L.-D., \& {Dai}, Z.-G. 2018, \apj, 861, 114

\bibitem[{{Yu} {et~al.}(2013){Yu}, {Zhang}, \& {Gao}}]{Yu2013}
{Yu}, Y.-W., {Zhang}, B., \& {Gao}, H. 2013, \apjl, 776, L40

\bibitem[{{Zhang} {et~al.}(2018){Zhang}, {Zhang}, {Sun}, {Lei}, {Gao}, {Li},
  {Shao}, {Zhao}, {Hu}, {L{\"u}}, {Wu}, {Fan}, {Wang}, {Castro-Tirado},
  {Zhang}, {Yu}, {Cao}, \& {Liang}}]{Zhang2018}
{Zhang}, B.~B., {Zhang}, B., {Sun}, H., {et~al.} 2018, Nature Communications,
  9, 447

\bibitem[{{Zhu} {et~al.}(2020){Zhu}, {Yang}, {Liu}, {Huang}, {Zhang}, {Li},
  {Yu}, \& {Gao}}]{zhu2020}
{Zhu}, J.-P., {Yang}, Y.-P., {Liu}, L.-D., {et~al.} 2020, \apj, 897, 20

\end{thebibliography}

\end{document}